\documentclass[fleqn,usenatbib]{mnras}

\usepackage{newtxtext,newtxmath}
\usepackage[T1]{fontenc}

\DeclareRobustCommand{\VAN}[3]{#2}
\let\VANthebibliography\thebibliography
\def\thebibliography{\DeclareRobustCommand{\VAN}[3]{##3}\VANthebibliography}

\usepackage{graphicx}	% Including figure files
\usepackage{amsmath}	% Advanced maths commands
\usepackage{microtype} % used to reduce spacing between text to fit in 5 pages

\usepackage{etoolbox, siunitx}
\robustify\bfseries

\usepackage{booktabs} % for tabular* environment

\usepackage[normalem]{ulem} % strikethrough text, e.g. \sout{insert text here}
\newcommand{\sophie}{SOPHIE}

\newcommand{\tess}{\emph{TESS}}

\newcommand{\pymcthree}{\textsc{pymc3}}
\newcommand{\lightkurve}{\textsc{lightkurve}}
\newcommand{\exoplanet}{\textsc{exoplanet}}
\newcommand{\starry}{\textsc{starry}}
\newcommand{\celerite}{\textsc{celerite}}

\newcommand{\toi}{TOI-1259}
\newcommand{\toip}{TOI-1259$\,$b}

\DeclareSIUnit\au{AU}
\DeclareSIUnit\Rsun{R_\odot}
\DeclareSIUnit\Rjup{R_\text{Jup}}
\DeclareSIUnit\Msun{M_\odot}
\DeclareSIUnit\Mjup{M_\text{Jup}}
\DeclareSIUnit\gyr{Gyr}
\DeclareSIUnit\ppt{ppt}
\DeclareSIUnit\ppm{ppm}

\newcommand*{\ra}[2][]{{% extra pair of braces to keep the \def-intion local!
    \def\SIUnitSymbolDegree{\textsuperscript{h}}%
    \def\SIUnitSymbolArcminute{\textsuperscript{m}}%
    \def\SIUnitSymbolArcsecond{\textsuperscript{s}}%
    \ang[#1]{#2}
    }%
}
\title[TOI-1259Ab -- exoplanet with white dwarf]{TOI-1259Ab -- a gas giant planet with 2.7\% deep transits and a bound white dwarf companion\thanks{Based in part on observations made at Observatoire de Haute Provence (CNRS), France.}}

\author[Martin et al.]{%
        David V. Martin$^{1}$\thanks{\href{mailto:martin.4096@osu.edu}{martin.4096@osu.edu}}\thanks{Fellow of the Swiss National Science Foundation},
        Kareem El-Badry$^{2}$,
        Vedad Kunovac Hod\v{z}i\'c$^{3}$\thanks{Fulbright Fellow},\newauthor
        Amaury H.\,M.\,J. Triaud$^{3}$,
        Ruth Angus$^{4,5}$,
        Jessica Birky$^{6}$,
        Daniel Foreman-Mackey$^{5}$,\newauthor
        Christina Hedges$^{7,19}$,
        Benjamin T. Montet$^{8}$,
        Simon J. Murphy$^{9}$, 
        Alexandre Santerne$^{10}$,\newauthor
        Keivan G. Stassun$^{11}$,
        Alexander P. Stephan$^{1,12}$,
        Ji Wang$^{1}$,
        Paul Benni$^{13}$,\newauthor
        Vadim Krushinsky$^{14}$,
        Nikita Chazov$^{15}$,
        Nikolay Mishevskiy$^{16}$,
        Carl Ziegler$^{17}$,\newauthor
        Abderahmane Soubkiou$^{18}$,
        Zouhair Benkhaldoun$^{18}$,
        Isabelle Boisse$^{10}$,
        Matthew Battley$^{29,30}$,\newauthor
        Nicola J. Miller$^{30}$,
        Douglas A. Caldwell$^{19,20}$,
        Karen Collins$^{20}$,
        Christopher E. Henze$^{19}$,\newauthor
        Natalia M. Guerrero$^{22}$,
        Jon M. Jenkins$^{19}$,
        David W. Latham$^{21}$,
        Adam Levine$^{22}$,\newauthor
        Scott McDermott$^{23}$,
        Susan E. Mullally$^{24}$, 
        George Ricker$^{22}$, 
        Sara Seager$^{22,25,26}$,\newauthor
        Avi Shporer$^{22}$,
        Andrew Vanderburg$^{27}$,
        Roland Vanderspek$^{22}$,
        Joshua N.\ Winn$^{28}$ \newauthor\emph{\normalsize Affiliations are listed at the end of the paper}
}

\date{Under review at MNRAS}

\pubyear{2021}

\begin{document}

\label{firstpage}
\pagerange{\pageref{firstpage}--\pageref{lastpage}}
\maketitle

\begin{abstract}
%We present TOI-1259Ab, a gas giant transiting a K dwarf on a 3.48 day orbit, with a bound white
    We present TOI-1259Ab, a $1.0R_{\rm Jup}$ gas giant planet transiting a $0.71R_{\odot}$ K-dwarf on a 3.48 day orbit. The system also contains a bound white dwarf companion TOI-1259B with a projected distance of $\sim1600$ AU from the planet host. Transits are observed in nine TESS sectors and are 2.7\% deep -- among the deepest known -- making TOI-1259Ab a promising  target for atmospheric characterization. Our follow-up radial velocity measurements indicate a variability of semiamplitude $K=71\,\rm m\,s^{-1}$, implying a planet mass of $0.44M_{\rm Jup}$.  By fitting the spectral energy distribution of the white dwarf we derive a total age of $4.08^{+1.21}_{-0.53}$\,Gyr for the system. The K dwarf's light curve reveals rotational variability with a period of 28 days, which implies a gyrochronology age broadly consistent with the white dwarf's total age.

\end{abstract}


%%%%%%%%%%%%%%%%%%%%%%%%%%%%%%%%%%%%%%%%%%%%%%%%%%

%%%%%%%%%%%%%%%%% BODY OF PAPER %%%%%%%%%%%%%%%%%%

\section{INTRODUCTION}\label{sec:introduction}

We know that roughly half of the stars in the galaxy exist in multiples \citep{Duquennoy1991,tokovinin2014}, but the vast majority of exoplanet discoveries have been in single star systems. The presence of a stellar companion will affect exoplanet populations. It may restrict the regions where planets may orbit stably \citep{dvorak1984,holman1999,mardling2001}, reduce the lifetime of protoplanetary discs \citep{kraus2012,daemgen2015,cheetham2015}, inhibit planetesimal formation \citep{thibault2008,xie2009} and induce high-eccentricity dynamics \citep{mazeh1979,eggleton2006,fabrycky2007}. Early exoplanet searches avoided close stellar multiples, whereas more distant binary companions were often undetected. Now, the vast {\it Gaia} astrometry survey is revealing thousands of wide binaries \citep{elbadry2018,Hartman2020,mugrauer2020}, many of which contain confirmed or candidate planets. 

An even less studied aspect of exoplanet populations is the effect of stellar evolution, with most planets being discovered around main sequence stars. As stars evolve they will expand, lose mass and, in most cases, leave behind a degenerate white dwarf. A surprising discovery is that up to roughly $50\%$ of white dwarfs have atmospheres polluted with heavy elements \citep{debes2012,farihi2016,wilson2019}, despite the fact that the high gravity should cause such elements to settle out of the atmosphere in a short time. This is seen as evidence that circumstellar  planetary material occasionally accretes onto white dwarfs, replenishing the heavy elements. 

However, it is tricky to actually find planets around white dwarfs. A lack of sharp spectral features prevents precise radial velocity  (RV) monitoring \citep{maxted2000}. A small radius, similar to that of Earth, significantly reduces transit probabilities and durations \citep{farmer2003,faedi2010}. A typically faint apparent magnitude leads to noisy light curves.  Astrometric planet detection with {\it Gaia} is promising, but will still be challenging because of the faintness of the objects \citep{silvotti2014}. Evidence for circumbinary planets has been presented for some binaries containing at least one white dwarf (e.g. \citealt{qian2009}), but the validity of the evidence has been repeatedly questioned \citep{zorotovic2013,wittenmyer2013,bear2014}. Only recently did \citet{vanderburg2020} discover the first bona fide planet transiting a white dwarf: WD 1856+534 (see also \citealt{alonso2021}). There have also been discoveries of transiting planetary debris \citep{vandenberg2015,manser2019,Vanderbosch2020,guidry2020} and accretion onto a white dwarf attributed to the evaporating atmosphere of a giant planet \citep{gansicke2019}.

In this paper we present TOI-1259Ab, a planet that is relevant to questions at the intersection of stellar evolution and stellar multiplicity. It is a transiting Jupiter-sized planet on a 3.48-day orbit around a K-dwarf, with a white dwarf companion at a projected separation of $\approx1600$ AU. The white dwarf was already known to be bound based on its Gaia parallax and common proper motion \citep{elbadry2018}. The transits were discovered by the TESS Science Processing Operations Center \citep{jenkins2016} and the community was alerted by the TESS Science Office on 17 October 2019 (Guerrero 2020), but the unusually deep transits of 2.7\% raised concerns that the signal is actually due to an eclipsing binary. Through our  RV follow-up we confirm that the signal is due to a planet, with a mass of $0.44 M_{\rm Jup}$. We summarise the key aspects of the TOI-1259 system in Table~\ref{tab:summary}.

Only a few bona fide planets have been discovered with degenerate outer companions (Table~\ref{tab:known_planets}), the first being Gliese-86b \citep{queloz2000,els2001,lagrange2006}. \citet{mugrauer2019} found 204 binary companions in a sample of roughly 1300 exoplanet hosts, of which eight of the companions were white dwarfs. \citet{mugrauer2020} found five white dwarf companions to TESS Objects of Interest, including TOI-1259, but without  RV data to confirm the TOIs as planets. Some of these planets were also in the \citet{elbadry2018} catalogue.

Even when a planet host star is still on the main sequence, the evolution of an outer companion still has implications for the planet's dynamics and survival \citep{kratter2012,stephan2020}. Heavy element pollution in any of these white dwarfs may be caused by its stellar binary companion \citep{veras2011,veras2013,bonsor2015,hamers2016,stephan2017}, and would also suggest both stars in the binary host (or once hosted) planets, of which only two systems are presently known (WASP-94, \citealt{neveuvanmalle2014} and XO-2, \citealt{Desidera14}). The presence of a white dwarf companion also makes it possible to calculate the system's age independently of other methods such as gyrochronology and isochrone fitting \citep{barnes2003,jorgensen2005,angus2019}.

\begin{figure}%[htpb]
    \centering
    \includegraphics[width=\linewidth]{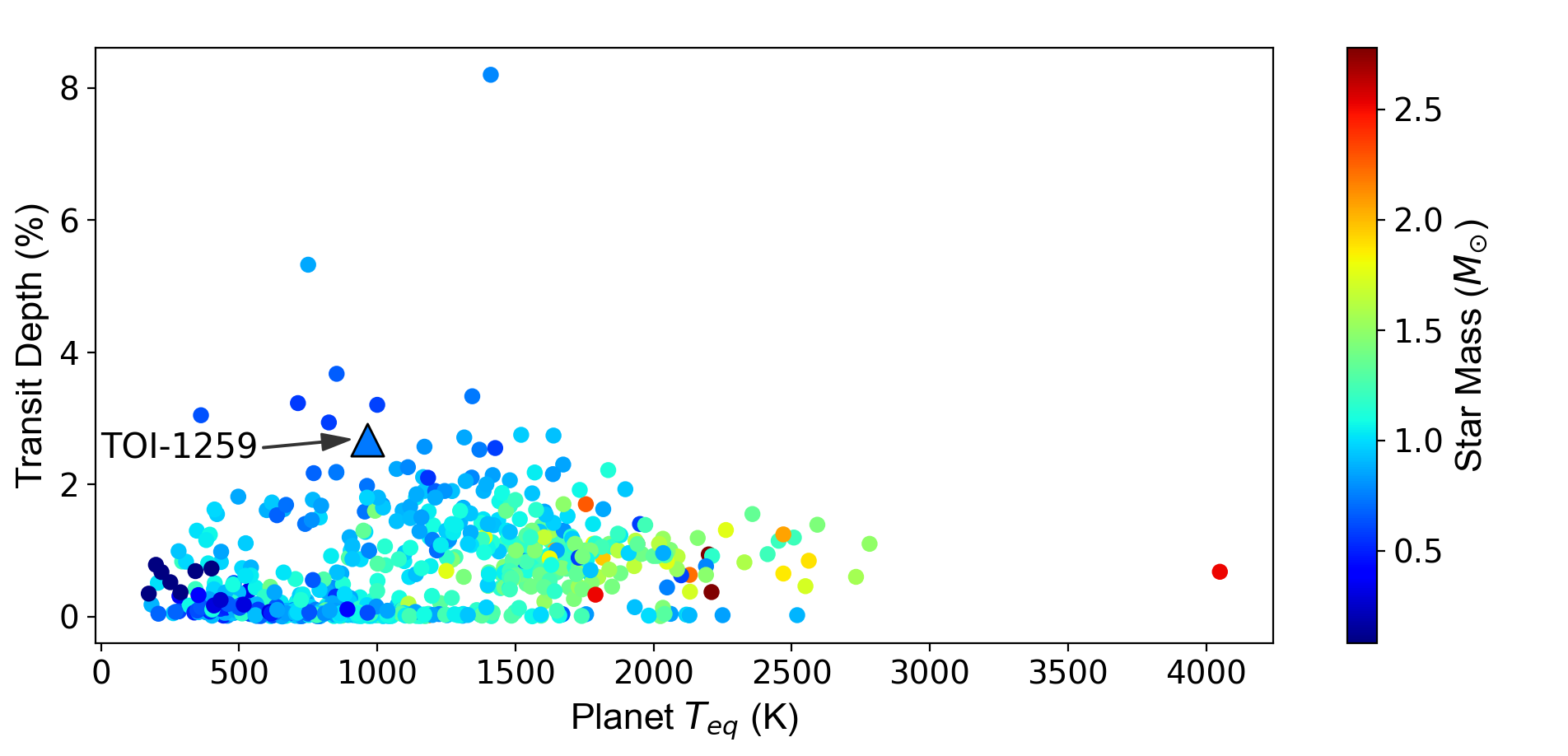}
    \caption{Transit depths of all confirmed transiting exoplanets as a function of the planet's equilibrium temperature (where such a value has been calculated). The colour indicates the mass of the host star in solar masses. The triangle demarcates TOI-1259 with $2.7\%$ deep transits.}
    \label{fig:transit_depths}
\end{figure}

White dwarf aside, the planet TOI-1259Ab has some beneficial properties for future atmospheric follow-up with JWST.  The planet has 2.7\% deep transits on its 0.71$R_{\odot}$ K-dwarf host, which are amongst the deepest known (Fig.~\ref{fig:transit_depths}). The $0.71R_{\odot}$ K-dwarf host star has a J magnitude of 10.226, and is located on the sky with an ecliptic latitude of $76.878^{\circ}$, placing it near the TESS and JWST continuous viewing zones. A measurement of the planet's atmospheric composition or other properties would also complement any measurement of pollution in the white dwarf atmosphere, as we try to better understand the formation and survival of planets in multi-stellar systems.

Our paper is structured as follows. Sect.~\ref{sec:observations} details the TESS photometry, SOPHIE  RVs and GAIA astrometry. In Sect.~\ref{sec:results} we present a combined analysis of these data and thereby characterise the planet, its host star and the companion white dwarf. We conclude by discussing some implications for this system and potential future work in Section~\ref{sec:discussion}.

\begin{table}
%    \renewcommand{\arraystretch}{1.2}
%    \sisetup{round-mode=places}

    \centering
    \caption{Summary of the TOI-1259 system. Host star parameters derived from SED fits (Sect.~\ref{subsec:sed}). White dwarf parameters are detailed in Table~\ref{tab:wd_fit}, and we only show parameters from the first model in that table here. Full planet parameters are shown in Table~\ref{table:results}. Coordinates and distances are from the TESS Input Catalog v8.1.}  %title of the table
%    \resizebox{\columnwidth}{!}{
    % align at dash later
    % https://tex.stackexchange.com/questions/4964/align-equals-sign-in-table
    \footnotesize
    \begin{tabular*}{\columnwidth}{@{\extracolsep{\fill}}
        % l
        l
        l
        l
        }
        \toprule
        \toprule
        Parameter & Description & Value \\
        \midrule

        \multicolumn{3}{@{}l@{}}{\emph{Host star -- TOI-1259A}} \\[10pt]
        ${\rm TIC}$ & TESS Input Catalog & 288735205 \\
        ${\rm GAIA}$ & ID & 2294170838587572736 \\
        $\alpha$ & Right ascension & $282.100297136879^{\circ}$  \\
        & & (\ra[angle-symbol-over-decimal,minimum-integer-digits=2]{18;48;24.07}) \\
        $\delta$ & Declination & $79.2560447193138^{\circ}$ \\
        & & (\ang[angle-symbol-over-decimal]{+79;15;21.76}) \\
        % $V_\mathrm{mag}$ & Apparent magnitude & 11.975 \\[5pt]
        $V_\mathrm{mag}$ & Apparent V magnitude & 12.08 \\
        $d$  & Distance (pc) & 118.11$\pm$0.37 \\
        $M_\star$  & Mass (\si{\Msun}) & $0.68^{+0.10}_{-0.01}$ \\
        $R_\star$  & Radius (\si{\Rsun}) & $0.739\pm 0.031$ \\
        $T_{\rm eff,\star}$  & Effective temperature (K) & 4775$\pm$100 \\
        ${\rm [Fe/H]}_{\star}$  & Metallicity & $-0.5\pm 0.5$ \\
        $\log g_{\star}$  & Surface gravity (cgs) & $4.5 \pm 0.5$ \\[10pt]
        
        \hline
        \multicolumn{3}{@{}l@{}}{\emph{Transiting planet -- TOI-1259Ab}} \\[10pt]
        $M_\text{pl}$  & Mass (\si{\Mjup}) & $0.441^{+0.049}_{-0.047}$ \\
        $R_\text{pl}$  & Radius (\si{\Rjup}) & $1.022^{+0.030}_{-0.027}$ \\
        $P_{\rm pl}$  & Orbital period (\si{\day}) & $3.4779780^{+0.0000019}_{-0.0000017}$ \\
        $a_\text{pl}$  & Semi-major axis (\si{\au})& $0.04070^{+0.00114}_{-0.00110}$ \\
        $e_\text{pl}$  & Eccentricity & $0$ \\[10pt]
        
        \hline
        \multicolumn{3}{@{}l@{}}{\emph{Bound white dwarf companion -- TOI-1259B}} \\[10pt]
        ${\rm TIC}$ & TESS Input Catalog & 1718312312 \\
        $\rm {GAIA}$ & ID & 2294170834291960832 \\
        $\alpha$ & Right ascension & $282.11052432749^{\circ}$ \\
        & & (\ra[angle-symbol-over-decimal,minimum-integer-digits=2]{18;48;26.53}) \\
        $\delta$ & Declination & $79.2594025546024^{\circ}$ \\
        & & (\ang[angle-symbol-over-decimal]{+79;15;33.85}) \\
        % $V_\mathrm{mag}$ & Apparent magnitude & 11.975 \\[5pt]
        $V_\mathrm{mag}$ & Apparent V magnitude & 19.23 \\
        $d$ & Distance  (pc) & 120.6$\pm$4.6 \\
        $M_{\rm WD}$  & Mass (\si{\Msun}) & $0.561\pm0.021$ \\
        $R_{\rm WD}$  & Radius (\si{\Rsun}) & 0.0131$\pm$0.0003 \\
        ${\rm sep}_\text{WD}$  & Projected current separation (\si{\au})& 1648 \\
        $T_{\rm eff,WD}$  & Effective temperature (\si{\kelvin}) & $6300^{80}_{-70}$ \\

        \bottomrule
    \end{tabular*}
    \label{tab:summary}
\end{table}

\begin{table*}
\caption{Known extra-solar planets around a main sequence star with a bound white dwarf companion, ordered by the current projected separation of the white dwarf (sep$_{\rm WD}$). Of the TOIs (TESS Objects of Interest) with white dwarf companions in the catalogues of \citet{elbadry2018} and \citet{mugrauer2020}, only TOI-1259Ab is confirmed to be a planet. CTOI-53309262 was only seen to transit once by TESS and so its period and semi-major axis are to be determined (TBD).}
\begin{tabular}{ccccc}
Name       & $a_{\rm pl}$ & ${\rm sep}_{\rm WD}$  & Planet Reference            & White Dwarf Reference                   \\
 &                 (AU)      &     (AU)            & &          \\
          \hline \hline
\begin{tabular}{@{}c@{}}HD 13445 \\  (Gliese-86) \end{tabular}  & 0.1143 & 21 & \citet{queloz2000} & \begin{tabular}{@{}c@{}}\citet{els2001} \\  \citet{lagrange2006}\end{tabular} \\\hline
\begin{tabular}{@{}c@{}}HD 27442 \\  (Epsilon Reticuli) \end{tabular}  & 1.271 & 236 & \citet{butler2001} & \begin{tabular}{@{}c@{}}\citet{chauvin2006} \\ \citet{mugrauer2007}\end{tabular} \\\hline 
HIP 116454  & 0.098 & 524 & \citet{vanderburg2015a} & \citet{vanderburg2015a} \\\hline
HD 8535 & 2.45 & 560 & \citet{naef2010} & \citet{mugrauer2019} \\ \hline
CTOI-53309262  & TBD & 625 & \begin{tabular}{@{}c@{}}Unconfirmed Community TOI \\  (Single Transit Only) \end{tabular} &\begin{tabular}{@{}c@{}}\citet{elbadry2018} \\  \citet{mugrauer2020} \end{tabular} \\\hline
Kepler-779  & 0.0558 & 1105 & \citet{morton2016} &\citet{mugrauer2019} \\\hline
TOI-1703  & 0.0223 & 1302 & Unconfirmed TOI &\citet{mugrauer2020} \\\hline
TOI-1259  & 0.0416 & 1648 & This Paper &\begin{tabular}{@{}c@{}}\citet{elbadry2018} \\  \citet{mugrauer2020} \end{tabular} \\\hline
HD 107148 & 0.269 & 1790 & \citet{butler2006} & \citet{mugrauer2016} \\\hline
TOI-249  & 0.0564 & 2615 & Unconfirmed TOI &\begin{tabular}{@{}c@{}}\citet{elbadry2018} \\  \citet{mugrauer2020} \end{tabular} \\\hline
WASP-98                 & 0.0453 & 3500 & \citet{hellier2014} & \begin{tabular}{@{}c@{}}\citet{mugrauer2019} \\  \citet{southworth2020}\end{tabular} \\\hline
HD 118904  & 1.7 & 3948 & \citet{jeong2018} & \citet{mugrauer2019} \\\hline
TOI-1624  & 0.0688 & 4965 & Unconfirmed TOI &\citet{mugrauer2020} \\\hline
\begin{tabular}{@{}c@{}}HD 147513 \\  (62 G. Scorpii) \end{tabular} & 1.32 & 5360 & \citet{mayor2004} & \citet{alexander1969} \\\hline

\end{tabular}

\label{tab:known_planets}

\end{table*}

\section{Observations}\label{sec:observations}

\subsection{\tess{} photometry}\label{subsec:observations_tess}

% We analyse \num{240} days of short cadence \tess{} photometry from sectors 14, 17--21, and 24--26  observed between 18 July 2019 and 4 July 2020. 
The \tess{} mission observed \toi{} in 2-minute cadence mode for a total of \num{240} days, covering nine sectors (14, 17--21, 24--26) between 18 July 2019 and 4 July 2020. The \tess{} Science Processing Operations Center (SPOC) pipeline \citep{jenkins2016} identified a \SI{\sim 2.5}{\percent} transit signal lasting \SI{\sim 2.2}{\hour}, repeating with a period of \SI{3.48}{\day}. The flat-bottomed shape and deep transit -- combined with the K dwarf host -- is consistent with a giant planet, substellar object, or very low-mass star.  There were 58 transit events in the \tess\ data.

%We downloaded the \tess{} 2-minute cadence SAP light curves  for \toi{} from sectors 14, 17--21, 24--26 using the \lightkurve{} package \citep{lightcurve2018}. Aside for the transits, the SAP light curves are dominated by a \SIrange{1}{2}{\percent} periodic signal consistent with one or several spots that are carried around the star due to rotation (top panel in Fig.~\ref{fig:tess_fit}). 

%We downloaded the pre-search data conditioned (PDCSAP) 2-minute cadence lightcurves created by the SPOC pipeline \citep{Stumpe2012,smith2012,Stumpe2014,jenkins2016}. 
% Second, in removing the spot modulation the PDCSAP lightcurves have introduced some low frequency systematics that differ from sector to sector. For these reasons, we use only the SAP light curves and apply our own Gaussian processes to clean them.

The \tess{} spacecraft fires its thrusters to unload angular momentum from its reaction wheels every few days,  which may cause the images obtained in the timestamps to appear disjoint. To make sure the momentum dumps do not affect our further analysis, we identify the times of these events from the Data Quality Flags in the FITS files, and exclude the data obtained within four hours on either side of the thruster events.
% Since some sectors show variability timescales that differ from others, we treat each sector independently using a Gaussian process model for the correlated noise (see  Section~\ref{subsec:results_fit}).

% Seven sectors of short cadence TESS photometry in Fig.~\ref{fig:tess_sectors}.

% Phase-folded primary transit in Fig.~\ref{fig:primary_transit}.

% Phase-folded secondary transit in Fig.~\ref{fig:secondary_transit}.

\subsection{Ground based follow-up photometry}\label{subsec:observations_ground}

We acquired ground-based time-series follow-up photometry of TOI-1259A as part of the TESS Follow-up Observing Program (TFOP)\footnote{https://tess.mit.edu/followup}. A full transit was observed on UTC 2019 October 10 using an unfiltered diffuser and again on UTC 2019 October 17 in $g'$-band from the Deep Sky West 0.5-m telescope near Rowe, New Mexico, USA. An egress was observed on UTC 2019 October 6 in R- and V-band using two 0.4-m telescopes at Kourovka observatory of Ural Federal University near Yekaterinburg, Russia. A full transit was observed on UTC 2019 October 24 in $z'$-band from the 0.36-m telescope at Acton Sky Portal private observatory in Acton, MA, USA. A full transit was observed on UTC 2019 December 4 in R-band from the 0.25-m telescope at Ananjev L33 private observatory near Ananjev, Ukraine. All observations detected on-time transits with depths consistent with TESS using apertures that were not blended with any known TICv8 or Gaia DR2 neighboring stars, except the diffuser observation was partially contaminated with a star that is too faint to cause the transit detection. Although \tess{} observed 58 transits of \toip{} in Cycle 2, we also include all available ground-based light curves in our analysis, with the exception of the $R$- and $V$-band observations at Kourovka Observatory on 2019 October 6 which only observed the egress. The reduced data of all ground-based photometry are available at ExoFOP-TESS\footnote{https://exofop.ipac.caltech.edu/tess}.
% Given the large number of TESS sectors available we choose for consistency to exclude the ground-based data in the model fits, but the reduced data are available at ExoFOP-TESS\footnote{https://exofop.ipac.caltech.edu/tess}.

\subsection{\sophie{} radial velocities}\label{subsec:observations_sophie}

To measure the mass of TOI-1259Ab we used SOPHIE, which is a high resolution\footnote{We used the ``high efficiency mode'', which has a resolution of $R=39000$ and is typically used for stars fainter than 10th magnitude.} \'echelle spectrograph used to find extra-solar planets with high-precision  RVs \citep{perruchot2008,bouchy2009}. It is installed on the 1.93 metre telescope at Observatoire de Haute Provence, France. 

We obtained 19 RV measurements of TOI-1259Ab between June 10 2020 and July 16 2020 using the SOPHIE spectrograph. All RVs were calculated using the standard SOPHIE pipeline, where a Cross-Correlation Function (CCF) is calculated between the data and a K5 mask. Each observation yielded an RV with a precision of roughly 20$\rm m\,s^{-1}$. 

We note that our spectra are not contaminated by the bound white dwarf, since it is both too faint (19.23 magnitude compared with 12.08 for the host star) and too far away (13.9 arcsec separation compared with 3 arcsec diameter SOPHIE fibres) to contribute
any light.

\subsection{Gaia Astrometry}\label{subsec:observations_gaia}

\begin{figure}%[htpb]
    \centering
    \includegraphics[width=\linewidth]{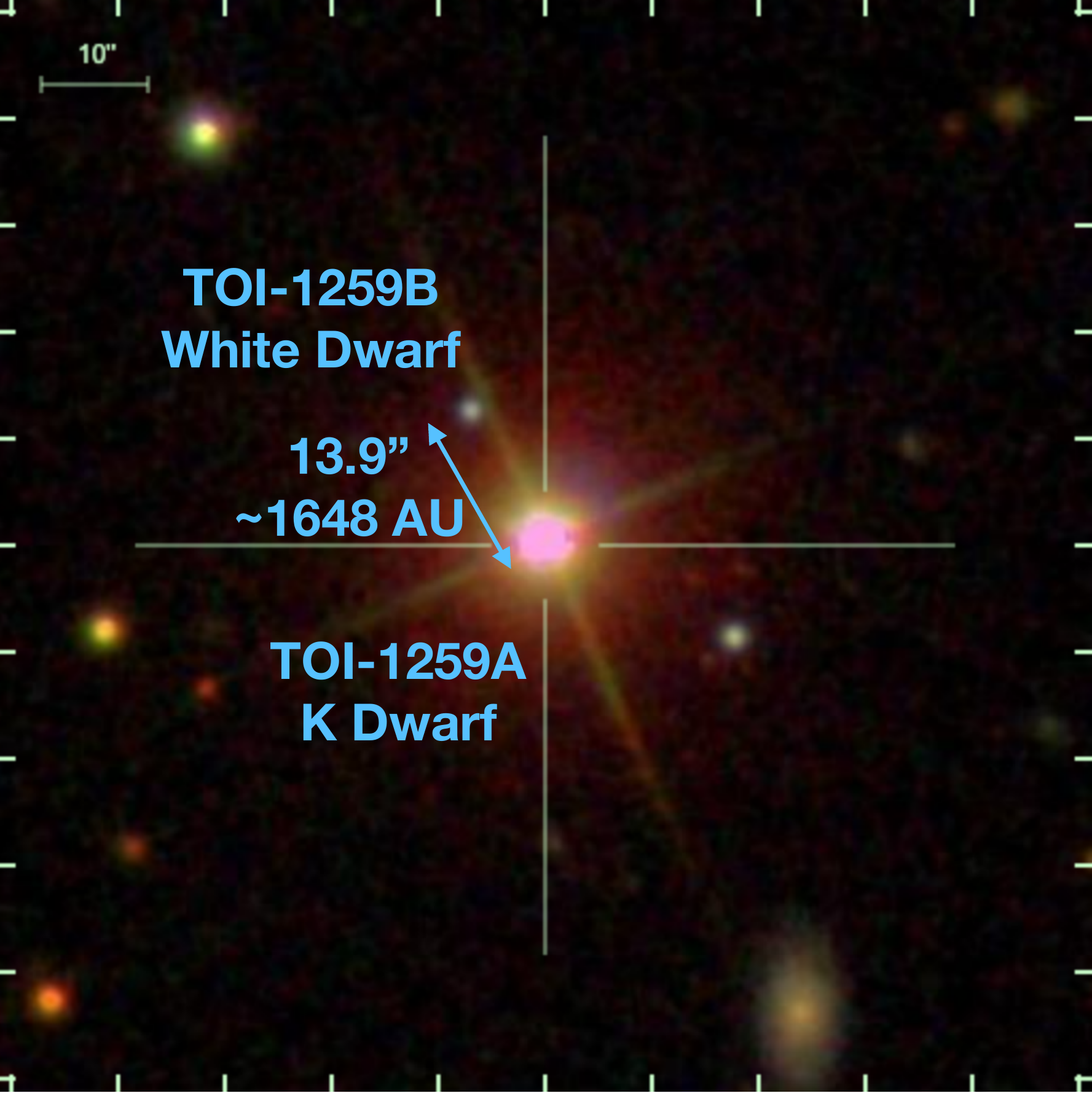}
    \caption{SDSS image of the planet host TOI-1259A and its bound white dwarf companion TOI-1259B. The image is centered on TOI-1259A at RA \ra[angle-symbol-over-decimal,minimum-integer-digits=2]{18;48;24.07}, Dec \ang[angle-symbol-over-decimal]{+79;15;21.76}.} 
    \label{fig:skymap}
\end{figure}

%{\bf SDSS or PanSTARRS image might be better? See e.g. \href{http://skyserver.sdss.org/dr14/en/tools/chart/navi.aspx?ra=282.11038931422&dec=79.25983609394&scale=0.1&width=60&height=60&opt=}{here} (centered on the WD)}.

%Very quick summary of why we are confident it is a bound system, based on El-Badry \& Rix 2018.

TOI-1259A and its white dwarf companion (TOI-1259B) were identified as a candidate wide binary by \citet{elbadry2018}, who searched {\it Gaia} DR2 for pairs of stars with positions, parallaxes, and proper motions consistent with bound Keplerian orbits\footnote{Orbital motion means that even bound orbits may have slightly different proper motions, but we require that this difference is less than the expected maximum orbital velocity at this separation.}.  Figure~\ref{fig:skymap} shows an SDSS image of the system, with both the K-dwarf primary and white dwarf companion clearly visible. The projected angular (physical) separation of the pair is 13.9 arcsec (1648 AU). The plane-of-the-sky absolute velocity difference between the WD and K dwarf is $\Delta V_{\perp} = 0.47^{+0.59}_{-0.25}\,\,{\rm km\,s^{-1}}$. For comparison, we can calculate the Keplerian orbital velocity of a circular orbit separated by 1648 AU, with masses $0.68$ and $0.561M_\odot$ (see Table~\ref{tab:summary}):

\begin{equation}
    V=\sqrt{\frac{G\left(M_\star + M_{\rm WD}\right)}{{\rm sep}_{\rm WD}}},
\end{equation}
where $G=6.67384\times10^{-11}$. The value $V=0.82{\rm km\,s^{-1}}$ is consistent with the {\it Gaia} measurement, within the precision of the measurements which are limited by the {\it Gaia} proper motion uncertainties. The semi-major axis of the WD - K dwarf orbit and the 3D separation of the two stars are not currently measurable. However, for randomly oriented orbits and a plausible eccentricity distribution, the projected semi-major axis is almost always within a factor of two of the true semi-major axis (see \citealt{elbadry2018}, their Figure B1). The white dwarf companion was later independently identified by \citet{mugrauer2020}.

Because orbital accelerations are not easy to measure in long-period binaries, distinguishing gravitationally bound wide binaries from chance alignments depends on statistical arguments about the probability of chance alignments. The chance alignment probability for a given separation, data quality, and background source density can be estimated empirically \citep[e.g.][]{lepine2007, elbadry2019, tian2020}. We follow the approach described in \citet{tian2020} to estimate the chance-alignment probability. In brief, we repeat the binary search after artificially shifting each star in {\it Gaia} DR2 by $\sim$1 degree, searching for companions around its new position. This procedure removes genuine binaries, but preserves chance alignment statistics (see \citealt{lepine2007}). Comparing the number of binary candidates found  so far to the number found in {\it Gaia} DR2 at similar separation, we estimate a chance-alignment probability of $\sim 1\times10^{-4}$ for TOI-1259A and its companion. That is, there is little doubt that the planet host and white dwarf are physically associated. 

\section{Analysis and results}\label{sec:results}
\subsection{Host star parameters}\label{subsec:sed}

\begin{figure}
\centering
\includegraphics[width=\linewidth,trim=100 75 85 90,clip]{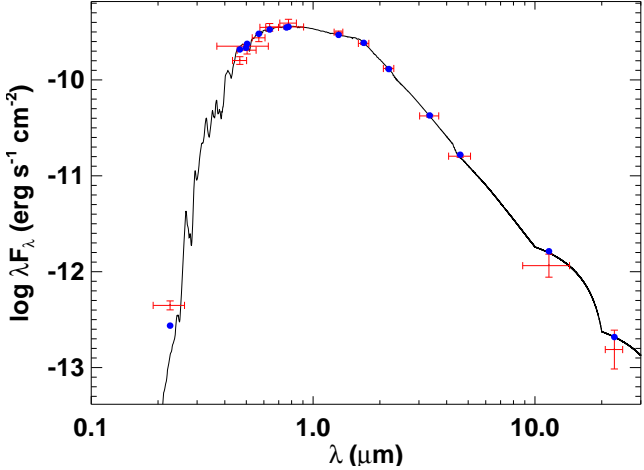}
\caption{Spectral energy distribution of TOI-1259A. Red symbols represent the observed photometric measurements, where the horizontal bars represent the effective width of the passband. Blue symbols are the model fluxes from the best-fit \citet{Kurucz1979} atmosphere model (black).  \label{fig:sed}}
\end{figure}

We performed an analysis of the broadband spectral energy distribution (SED) of the star together with the {\it Gaia\/} DR2 parallaxes \citep[adjusted by $+0.08$~mas to account for the systematic offset reported by][]{StassunTorres2018}, in order to determine an empirical measurement of the stellar radius, following the procedures described in \citet{Stassun2016,Stassun2017,Stassun2018}. We pulled the $BVgri$ magnitudes from {\it APASS}, the $JHK_S$ magnitudes from {\it 2MASS}, the W1--W4 magnitudes from {\it WISE}, the $G G_{\rm BP} G_{\rm RP}$ magnitudes from {\it Gaia}, and the NUV magnitude from {\it GALEX}. Together, the available photometry spans the full stellar SED over the wavelength range 0.2--22~$\mu$m (see Figure~\ref{fig:sed}).  

We performed a fit to the SED using \citet{Kurucz1979} stellar atmosphere models, with the effective temperature ($T_{\rm eff}$), metallicity ([Fe/H]), surface gravity ($\log g$) as free parameters. The only additional free parameter is the extinction ($A_V$), which we restricted to the maximum line-of-sight value from the dust maps of \citet{schlegel1998}. The resulting fit is very good (Fig.~\ref{fig:sed}) with a reduced $\chi^2$ of 1.7 and best-fit $A_V = 0.20 \pm 0.07$, $T_{\rm eff} = 4775 \pm 100$~K, $\log g = 4.5 \pm 0.5$, and [Fe/H] $= -0.5 \pm 0.5$. Integrating the (unreddened) model SED gives the bolometric flux at Earth, $F_{\rm bol} = 5.94 \pm 0.14 \times 10^{-10}$ erg~s$^{-1}$~cm$^{-2}$. Taking the $F_{\rm bol}$ and $T_{\rm eff}$ together with the {\it Gaia\/} DR2 parallax, gives the stellar radius, $R_\star = 0.739 \pm 0.031$~R$_\odot$. 

 The stellar mass can be obtained from the SED analysis in two ways. First, we can use the $R_\star$ together with $\log g$ to obtain a mass estimate of $M_\star = 0.79 \pm 0.14$~M$_\odot$.  Alternatively, we can apply the \citet{Torres2010} empirical mass-radius relations to get a value $M_\star = 0.68 \pm 0.08$~M$_\odot$. 

 As an independent test of our stellar parameters, we ran \textsc{ExoFASTv2} \citep{Eastman2019} to create a joint fit of the SED and two different types of isochrones: MIST \citep{dotter2016,choi2016} and PARSEC \citep{Bressan2012}. With MIST we obtain $R_\star = 0.733\pm0.022$ and $M_\star=0.777^{+0.037}_{-0.038}$ and with PARSEC we obtain similar values of $R_\star = 0.729\pm0.020$ and $M_\star=0.777^{+0.034}_{-0.031}$.

 Throughout this paper we will use the $R_\star = 0.739R_\odot$ and $M_\star = 0.68 M_{\odot}$ values calculated using the empirical SED and \citet{Torres2010} relation, respectively. These are the priors that will be used in the global analysis to determine the planet parameters in Sect.~\ref{subsec:results_fit}.

\renewcommand{\arraystretch}{1.5}
\begin{table}
%    \renewcommand{\arraystretch}{1.2}
%    \sisetup{round-mode=places}

    \centering
    \caption{ Stellar radius and mass measurements based on four different methods. In all cases we use a fit to the SED combined with \textit{Gaia} DR2 parallaxes. In methods 1 and 2 we follow the procedure of \citet{Stassun2016,Stassun2017,Stassun2018} to derive the radius and the mass comes from the SED measurement of the surface gravity $\log g$ (1) and the \citet{Torres2010} mass-radius relationship (2). In methods 3 and 4 we use ExoFASTv2 \citep{Eastman2019} to fit the SED and two different isochrones. We use method 2 (in bold) as the nominal value.}  %title of the table
%    \resizebox{\columnwidth}{!}{
    % align at dash later
    % https://tex.stackexchange.com/questions/4964/align-equals-sign-in-table
    \footnotesize
    \begin{tabular*}{\columnwidth}{@{\extracolsep{\fill}}
        % l
        l
        l
        l
        }
        \toprule
        \toprule
        Method & Radius  & Mass \\
         & ($R_\odot$)  & ($M_\odot$) \\
        \midrule

        1. SED + $\log g$ & $0.739\pm0.031$ & $0.79\pm0.14$ \\
        {\bf 2. SED + \citet{Torres2010} M-R} & $0.739\pm0.031$ & $0.68\pm0.08$ \\
        3. SED + MIST isochrones & $0.733\pm0.022$ & $0.777^{+0.037}_{-0.038}$ \\
        4. SED + PARSEC isochrones & $0.729\pm0.020$ & $0.777^{+0.034}_{-0.031}$ \\

        \bottomrule
    \end{tabular*}
    \label{tab:summary}
\end{table}

\subsection{Host star rotation}\label{subsec:results_host_rotation}

%The photometric rotation period measured from the Gaussian process model gives $P_\mathrm{rot} = 41.5^{+5.7}_{-7.1}\,$d, which is significantly different from the $\SI{\sim27}{\day}$ rotation period that is apparent from the light curve in the top panel of Fig.~\ref{fig:tess_fit}. We present a possible explanation for this discrepancy. A Lomb-Scargle periodogram of the \tess{} light curve shows a clear dominant peak around 27 days, and additional broad power that extends beyond 40 days. The cause of the broad peak may be due to various effects, such as evolving star spots, differential rotation due to spots at different latitudes, or long-term \tess{} spacecraft systematics. The Gaussian process model with which we model the rotation is parametrized by two harmonic oscillators with a fundamental frequency and its first harmonic, and therefore most of the power can be described by a \SI{\sim40}{\day} signal and its half period. 

%We note that adding another Gaussian process term to model any additional non-periodic signal was not able to constrain the rotation further. 

\begin{figure}%[htpb]
    \centering
    \includegraphics[width=\linewidth]{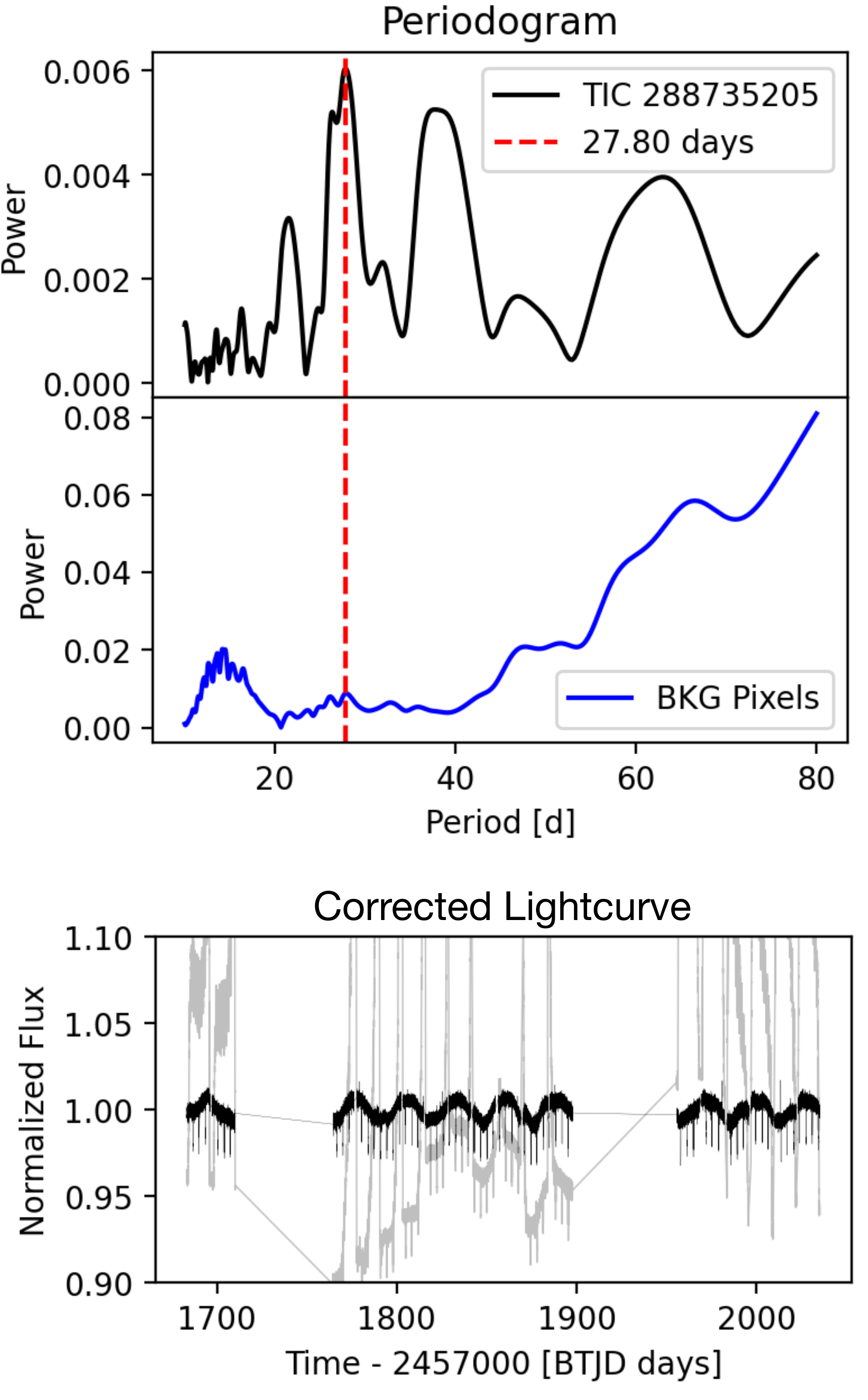}
    \caption{ Top: Systematics-Insensitive Periodogram (SIP) for TOI-1259b. The periodogram is calculated for both the corrected lightcurve (black line) and the background (BKG) pixels (blue line). There is a strong peak in the SIP at 28 days, which is attributed to the rotation of the planet host and denoted by a red dashed line. The background pixels show no evidence of any periodicity, suggesting that the 28 day signal is both real and intrinsic to the target. Note that these periodograms are not normalized by the measurement errors. Bottom: The light curve for TOI-1259. Grey points show the raw data TESS data, and black points show the data corrected using \texttt{tess-sip}, showing a clear periodicity.}
    \label{fig:sip}
\end{figure}

We use a Systematics-Insensitive Periodogram (SIP) to build a periodogram whilst simultaneously detrending TESS instrument systematics from scattered background, following the method first described in \cite{sip} and more recently implemented for TESS data in \citet{Hedges_2020}. For this we use the SAP lightcurves, since the PDCSAP lightcurves tend to have the rotation period removed, or at least made harder to identify. The SIP power amplitude is shown in Figure ~\ref{fig:sip}. The SIP shows the most significant power at a period of P$_{\rm rot}$=28 d and a secondary peak at P$_{\rm rot}\approx$40 days. We adopt the more significant peak at P$_{\rm rot}$=28 d as the true rotation rate of TOI-1259. The signal at P$_{\rm rot}\approx$40 days is possibly an alias of the rotation period.

The rotation period of P$_{\rm rot}$=28 d is close to the orbital period of TESS (27 days).  We conduct two tests of the validity of this rotation rate. First,  We construct SIPs for the targets neighboring TOI-1259 and find no evidence of a similar peak in near-by targets.  Second, we create a SIP for all “background” pixels outside of the TESS pipeline aperture in the TESS Target Pixel File for TOI-1259. This background SIP, shown as a blue line in Fig.~\ref{fig:sip}, has no power at 28 days. This suggests that the 28 day signal is intrinsic to the target, and not an artifact of, for example, the sampling frequency of TESS.

%Instead, improvements can be made in the light curve extraction by developing methods that preserve astrophysical variability in multi-sector \tess{} data. 

%However, this is beyond the scope of this paper. 

%We therefore urge caution using the rotation period from the Gaussian process, and instead adopt a rotation period of 27 days from the Lomb-Scargle periodogram.

% Moreover, the flux residuals exhibit a variation in phase with the orbit, which is plausible due to the intense irradiation of the planet at periastron. Contrary to tidally-locked planets on circular orbits, the functional form of the phase curves of planets on eccentric orbits is not well defined. For one, eccentric planets might be rotating pseudo-synchronously \citep{hut1981}, and experience time-variable heating throughout their orbit. However, for low-to-moderately eccentric systems, reasonable fits to phase curves may be obtained by small adjustments to the circular case.

\subsection{Host star age}\label{subsec:results_host_age}

When estimating the ages of K dwarfs using gyrochronology, it is essential to account for `stalled magnetic braking'. Recent observations have revealed that rotational evolution is inhibited for middle-aged K dwarfs \citep{curtis2019, angus2020}. This stalled rotational evolution is thought to be caused by an internal redistribution of angular momentum \citep{spada2020}. Unless this phenomenon is taken into account, the ages of K dwarfs could be underestimated by more than 2 Gyr.

We estimated an age for this star using a new gyrochronology model that accounts for stalled magnetic braking (Angus et al., in prep). This model was calibrated by fitting a Gaussian process, a semi-parametric model that is flexible enough to capture the complex nature of stellar spin-down, to a number of asteroseismic stars and open clusters, including NGC 6811 where many member stars exhibit stalled magnetic braking \citep{curtis2019}. We also used kinematic ages of {\it Kepler} field stars to calibrate this model for old K and early M dwarfs, where there is a dearth of suitable open cluster calibration stars. These kinematic ages also reflect the stalled magnetic braking behaviour seen in open clusters \citep{angus2020}.

Using this model, we infer an age of 4.8$^{+0.7}_{-0.8}$ Gyr for this star. The quoted age uncertainty is the formal uncertainty that results from the uncertainty on the star's rotation-period, and does not account for uncertainty in the model. Quantifying the magnitude of the model-uncertainty is beyond the scope of this paper, however a 20\% uncertainty of around 1 Gyr may be a more reasonable estimate of the true age-uncertainty. This estimated age for the planet host is consistent with that of the total age of the white dwarf (Sect.~\ref{subsec:results_white_dwarf_age}).

\subsection{White dwarf age}\label{subsec:results_white_dwarf_age}

White dwarfs (WDs) steadily cool as they age. A WD's cooling age -- that is, the time since it became a WD -- can therefore be constrained from its temperature and luminosity. If the WD's mass is known, the initial mass of its progenitor star can be inferred through the initial-final mass relation (IFMR), and this initial mass constrains the pre-WD age of the WD progenitor. Therfore if we have a well-constrained distance to the WD then its total age, i.e. the sum of its main sequence lifetime and its cooling age, can be robustly measured from its spectral energy distribution (SED). Under the reasonable ansatz that the WD and K dwarf formed at the same time, we can then measure the total system age from the WD. 

We use \texttt{BASE-9} \citep{vonhippel2006, degennaro2008, Stein2013, Stenning2016} to fit the SDSS $ugriz$ photometry of the the WD. \texttt{BASE-9} combines evolutionary models for WDs \citep{Althaus_1998, Montgomery1999}, WD atmospheric models \citep{Bergeron1995, Holberg2006}, PARSEC evolutionary models \citep{Bressan2012}, and semi-empirical IFMRs to predict the SED of a WD with a given age, initial mass and metallicity, distance, extinction, and spectral type. It then uses MCMC methods to constrain these parameters from the SED of an observed WD. 

We use the {\it Gaia} parallax of the brighter K dwarf companion as a prior. For the initial [Fe/H], we assume a Gaussian prior with a mean of -0.2 and a standard deviation of 0.3, appropriate for a disk star in the solar neighborhood \citep[e.g.][]{Hayden2015}. The spectral type of the WD is not known. Although we expect extinction to be almost negligible for such a nearby WD, we fit the extinction $A_V$ as a free parameter, with a Gaussian prior with a mean of 0.01 and a standard deviation of 0.02, based on the 3D dust map of \citet{Green2019}. For context, the SFD reddening (which quantifies the total dust column to infinity, including dust behind the WD) at the WD's position is $E(B-V) = 0.085$ \citep{schlegel1998}. We use a flat prior between 0 and 12 Gyr for total age.
Our fiducial fit assumes the WD has a hydrogen atmosphere, which is true for $\sim$75\% of WDs with its temperature and mass. We also show how the constraints would change if the WD has a helium atmosphere in Table~\ref{tab:wd_fit}. 

A systematic uncertainty in modeling the WD's evolution is the IFMR. Because most published IFMRs are discrete, analytic fitting functions, this uncertainty is difficult to marginalize over gracefully. To estimate the magnitude of this uncertainty, we compare constraints that assume two different IFMRs: the IFMR measured by \citet{williams2009} from bound clusters, and the IFRM measured by \citet{elbadry2018b} from the {\it Gaia} color-magnitude diagram of nearby field WDs.

\renewcommand{\arraystretch}{1.5}
\begin{table*}
	\centering
	\caption{ Parameters of the WD. We compare constraints derived assuming a hydrogen versus helium atmosphere, and constraints that assume two different IFMRs.}
	\label{tab:wd_fit}
	\begin{tabular}{lllll} 
		\hline
		Parameter & H atm; El-Badry+18 IFMR & H atm; Williams+09 IFMR & He atm; El-Badry+18 IFMR & He atm; Williams+09 IFMR \\
		\hline
		Total age (Gyr)      & $4.08^{+1.21}_{-0.53}$      &  $3.73^{+0.56}_{-0.31}$       & $4.67^{+1.70}_{-0.94}$       & $3.94^{+0.81}_{-0.44}$       \\
		Cooling age (Gyr)    & $1.88^{+0.07}_{-0.06}$      &  $1.88^{+0.07}_{-0.06}$       & $1.78^{+0.06}_{-0.05}$       & $1.78^{+0.06}_{-0.05}$        \\
		Pre-WD age (Gyr)     & $2.18^{+1.2733}_{-0.5597}$  &  $1.84^{+0.61}_{-0.35}$       & $2.87^{+1.73}_{-1.00}$       & $2.14^{+0.86}_{-0.47}$        \\
		Radius ($R_{\odot}$) & $0.0131^{+0.0003}_{-0.0003}$&  $0.0131^{+0.0003}_{-0.0003}$ & $0.0129^{+0.0003}_{-0.0003}$ & $0.0129^{+0.0003}_{-0.0003}$  \\
		Mass ($M_{\odot}$)   & $0.561^{+0.021}_{-0.021}$   &  $0.561^{+0.021}_{-0.021}$    & $0.548^{+0.021}_{-0.019}$    & $0.548^{+0.021}_{-0.019}$     \\
  Initial mass ($M_{\odot}$) & $1.59^{+0.22}_{-0.22}$      &  $1.72^{+0.17}_{-0.17}$       & $1.45^{+0.22}_{-0.20}$       & $1.61^{+0.17}_{-0.17}$        \\
	    $T_{\rm eff}$\,(K)   & $6300^{+80}_{-70}$      &  $6300^{+80}_{-70}$       & $6330^{+80}_{-70}$       & $6330^{+80}_{-70}$        \\
        $A_V$\,(mag)         & $0.019^{+0.018}_{-0.013}$   &  $0.019^{+0.018}_{-0.013}$    & $0.019^{+0.017}_{-0.013}$    & $0.019^{+0.017}_{-0.013}$     \\
		\hline
	\end{tabular}
\end{table*}

\begin{figure*}
    \centering
    \includegraphics[width=\textwidth]{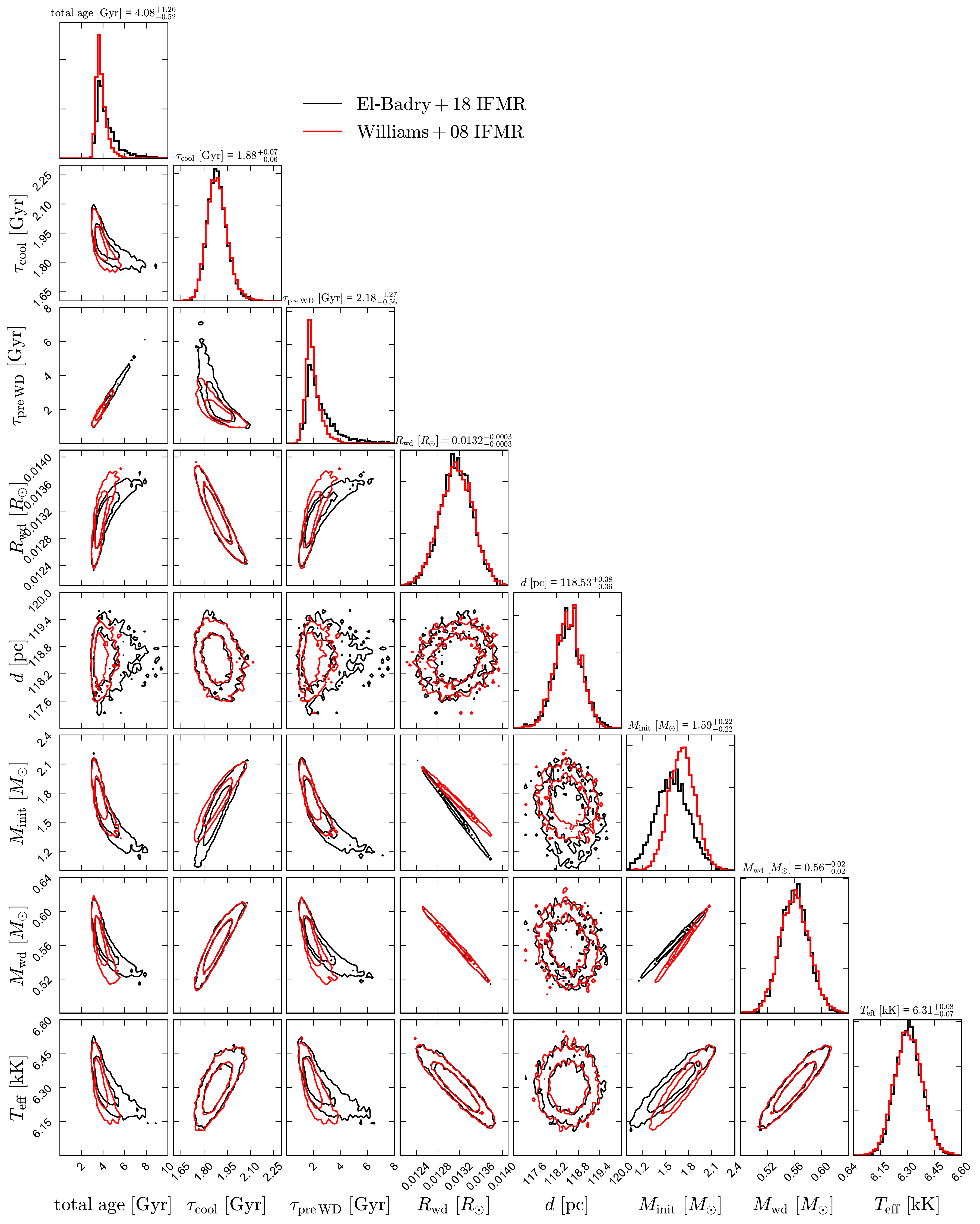}
    \caption{Parameters of the WD and its progenitor, obtained from fitting the SED. We compare constraints obtained when assuming the initial-final mass relations (IFMR) from \citet{elbadry2018b} and \citet{williams2009}. For this figure we assume a hydrogen atmosphere, but in Table~\ref{tab:wd_fit} we also show results for a potential helium atmosphere. Values listed on the diagonal are based on the \citet{elbadry2018b} IFMR. Contours enclose 68 and 95\% probability. The total age of the WD -- and thus, presumably, the age of the system -- is at least 3 Gyr. The age uncertainty is dominated by uncertainty in the mass and pre-WD age of the progenitor.}
    \label{fig:wd_fit}
\end{figure*}

Figure~\ref{fig:wd_fit} shows the resulting constraints on parameters of the WD, assuming a hydrogen atmosphere. Values are also reported in Table~\ref{tab:wd_fit}. The temperature and radius of the WD are well-constrained by the SED. Because the radius of a WD is determined primarily by its mass, this also constrains the WD's mass, which in turn constrains the initial mass of the WD progenitor. The cooling age of the WD is reasonably well-constrained to be between 1.7 and 2 Gyr. The pre-WD age is more uncertain, because a modest uncertainty in initial mass leads to a significant uncertainty in main-sequence lifetime. This is the primary cause of the differences in the constraints obtained for the two different IFMRs. The two relations are actually quite similar at the relevant WD mass (see \citealt{elbadry2018b}, their Figure 3), but the \citet{elbadry2018b} relation is somewhat shallower. This means that a larger range of initial masses could produce the observed WD mass, and thus, that there is a larger range of allowed pre-WD ages. The pre-WD lifetime of a $2 M_{\odot}$ star is $\approx 1.3$ Gyr, while that of a $1.2 M_{\odot}$ star is $\approx 6$ Gyr, so the resulting uncertainty is non-negligible. A tighter constraint on the WD mass -- which is potentially achievable via gravitational redshift; e.g. \citealt{Reid1996} -- could improve the statistical uncertainty on total age. However, the constraint is already tight enough that systematic uncertainty due to the IFMR is comparable to the statistical uncertainty, so the IFMR uncertainty would likely dominate the age uncertainty even with significantly better data.

Table~\ref{tab:wd_fit} also shows how constraints on the WD's parameters would change if it had a helium atmosphere rather than a hydrogen atmosphere. Changing the atmosphere slightly changes the WD's colors and the mass-radius relation, such that the implied mass of the WD is lower. The difference is, however, relatively modest. Obtaining a spectrum of the WD would remove this source of uncertainty, since a WD with a hydrogen atmosphere and $T_{\rm eff}> 6000$\,K would have detectable Balmer lines.  

Our derived total age for TOI-1259B is consistent with the 4.8$^{+0.7}_{-0.8}$ Gyr age derived for the planet host TOI-1259A based on gyrochronology (Sect.~\ref{subsec:results_host_age}).

\subsection{Radial velocity modelling}\label{subsec:results_rvfit}

 To confirm the planet we conduct two independent analyses. First, in this section we detect solely the  RV signal, using the genetic algorithm {\sc Yorbit} \citep{segransan10}. Second, in Sect.~\ref{subsec:results_fit} we do a combined fit of both the photometry and  RV, with a completely different code.

\begin{figure}%[htpb]
    \centering
    \includegraphics[width=0.99\linewidth]{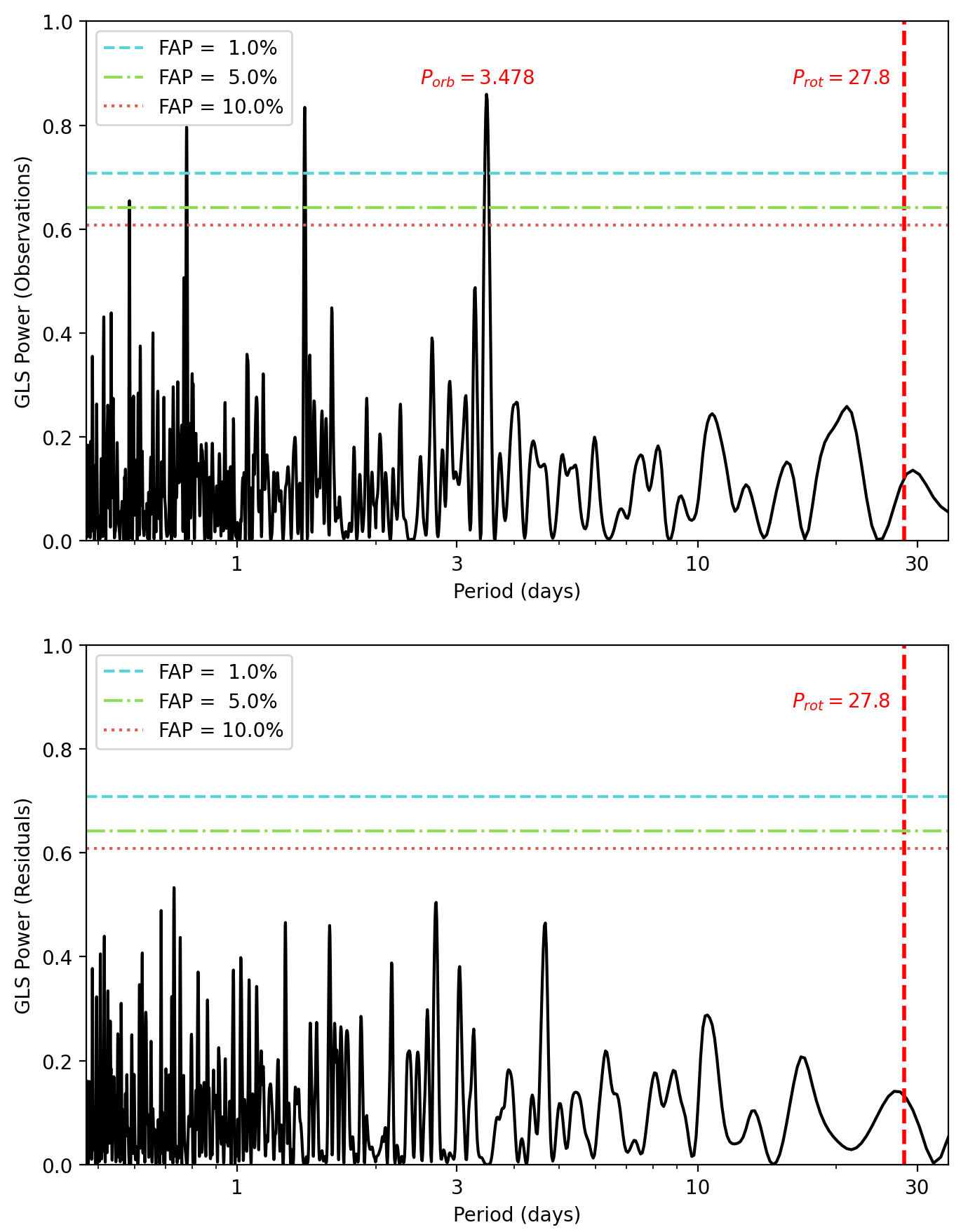}
    \caption{Generalized Lomb-Scargle (GLS) periodogram of the 19 SOPHIE  RV points  (top) and the residuals to the circular Keplerian fit (bottom). The  highest peak  for the observations corresponds to the transiting planet orbital period at 3.45 days.  There are also harmonics at integer fractions of this period. Note that these periodogram does not include any information from the photometry. The RV detection alone is  significant, above a 1\% false alarm propbability (FAP). For reference, we show the 27.8 day rotation period of the host star with a red dashed line, at which there is no power in the GLS. The GLS periodogram of the residuals shows no significant peak above a 10\% FAP.}
    \label{fig:periodogram}
\end{figure}

In Fig.~\ref{fig:periodogram} we show a  Generalized Lomb-Scargle periodogram derived solely from the SOPHIE RVs.  The 3.478 day period from the TOI catalogue is denoted by a red dashed line. This corresponds to the highest peak in the periodogram, demonstrating that the RV signal is produced by the same body that produces the TESS transit signals.  We note that our spectra are not contaminated by the bound white dwarf, since it is both too faint (19.23 magnitude compared with 12.08 for the host star) and too far away (13.9 arcsec separation compared with 3 arcsec diameter SOPHIE fibres) to contribute
any light.

 We then run the \textsc{Yorbit} genetic algorithm, allowing it to fit a single Keplerian model with a period between 3 and 4 days, which is roughly centered on the highest peak of the periodogram. The best-fitting model of the RVs alone has a period of $P_{\rm pl} = 3.42\pm0.06$ days, and a best-fitting transit mid point of $2037.96\pm0.14$ (BJD$_{\rm UTC}$ - 2 457 000), both of which match the TESS photometry. The best-fitting model also has an eccentricity of $e_{\rm pl } =0.178$, which would be surprisingly high for such a short period planet.

To test if the planet signal (and its potential eccentricity) is significant, we calculate the Bayesian Information Criterion ($BIC$) by 

\begin{equation}
    BIC = n\ln\left(\frac{RSS}{n}\right) + k\ln{n_{\rm obs}},
\end{equation}
where $k$ is the number of model parameters, $RSS$ is the sum of the squares of the model residuals (in $m/s$) and $n_{\rm obs} = 13$ is the number of observations. We calculate the BIC for a flat line ($k=1$), the best fitting eccentric model from \textsc{Yorbit} ($k=6$) and a forced circular model with the same period ($k=4$). The flat, eccentric and circular $BIC$ values are 157.8, 136.2 and 133.6, respectively. The circular planet model has the lowest $BIC$, making it the favoured model. For one model to be significantly better than another though, a $BIC$ reduction of more than 6 is considered ``strong evidence''. This means that the circular model is not significantly better than the eccentric model, but both are significantly better than the flat model. Otherwise said, the  RVs alone provide strong evidence that the planet exists, but we cannot constrain its eccentricity.

%In Fig.~\ref{fig:radial_velocities} we plot our radial velocities, along with Keplerian curves derived from our combined fit of the photometric and radial velocity data (described in Sect.~\ref{subsec:results_fit}). Our radial velocity measurements are all published online.

%Whilst the period and phase of the fit are tightly constrained
%by the TESS photometry, the RVs by themselves show strong evidence for
%the planet.  Using the genetic algorithm {\sc Yorbit} \citep{segransan10}, the best-fitting circular model of the RVs alone has a period of $P_{\rm pl} = 3.42\pm0.06$ days, and a best-fitting transit mid point of $2037.96\pm0.14$ (BJD$_{\rm UTC}$ - 2 457 000), both of which match the TESS photometry. In Fig.~\ref{fig:periodogram} we show a Lomb-Scargle periodogram derived solely from the SOPHIE RVs, which independently finds the planet's orbital period and a 2:1 harmonic. This demonstrates that the RV signal is produced by the same body that produces the TESS transit signals.  

%With a radial velocity peak-to-peak amplitude of less than 200ms$^{-1}$, it is clear that the transiting object is planetary in nature. We derive a mass of $0.429^{+0.040}_{-0.036}M_{\rm Jup}$. For this mass the Chen \& Kipping probabilistic mass-radius relation derives a radius of $1.25^{+0.32}_{-0.27}$, which is compatible with our measured value of $1.022^{0.030}_{-0.027}R_{\rm Jup}$ and hence there are no significant signs of inflation of the planet's radius. The planet's scaled separation is $a/R_\star=12.6$ and we deduce an equilibrium temperature of $T_{\rm eq}=936K$. 

\subsection{Global modelling of the photometry and radial velocity}\label{subsec:results_fit}

\begin{figure*}
    \centering
    \includegraphics{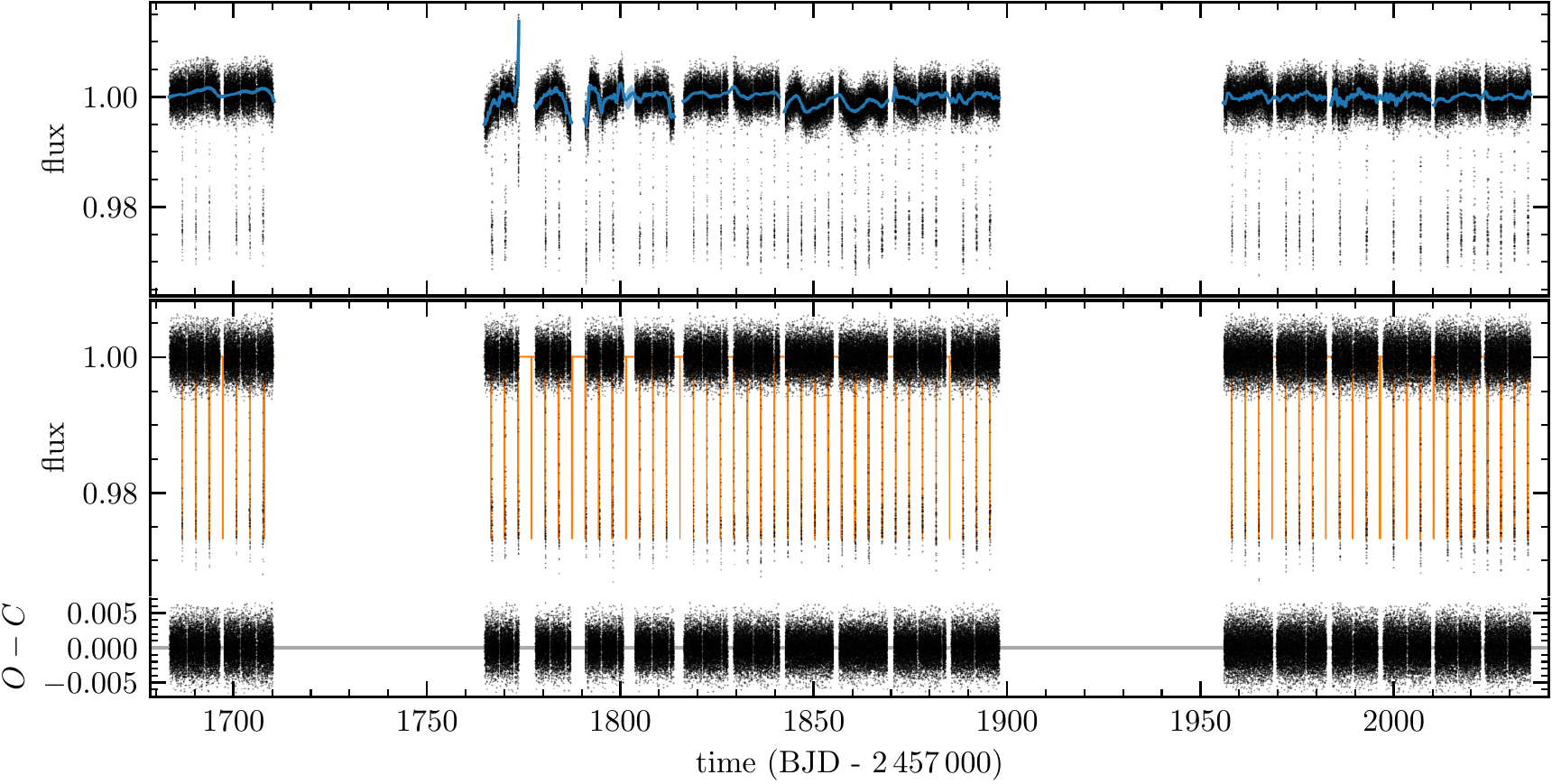}
    \caption{\emph{Upper panel:} Nine sectors of \tess{} PDCSAP photometry of \toi{}. The blue model is the Gaussian process model. \emph{Middle panel:} The light curve after removing the Gaussian process model, showing transits only, with the transit model overlaid in orange. \emph{Bottom panel:} Residuals from the best-fitting full model.}
    \label{fig:tess_fit}
\end{figure*}

\begin{figure}%[htpb]
    \centering
    \includegraphics[width=\linewidth]{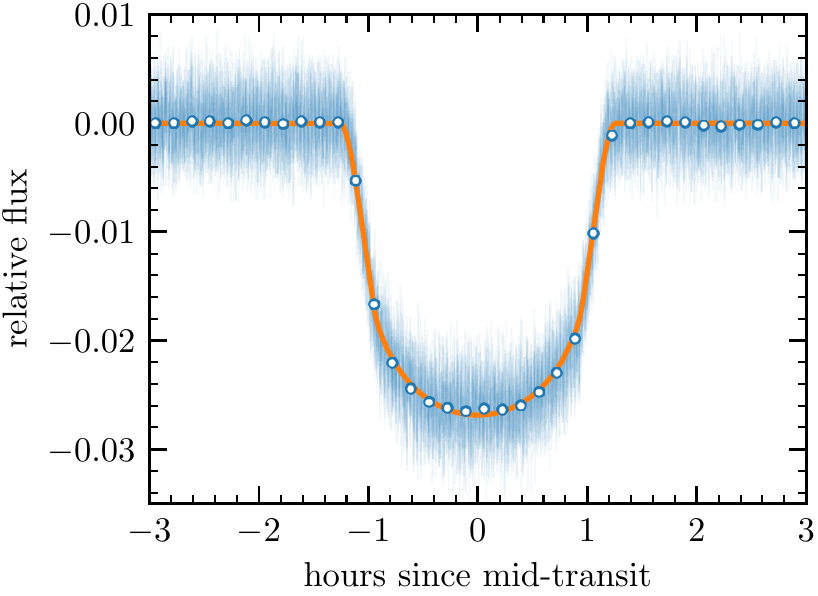}
    \caption{Phase folded and detrended light curve of the primary transit after removing the Gaussian process model. We show the unbinned data in blue, and the data averaged in \SI{10}{\minute} bins in blue/white points. The orange line is the best-fitting transit model, with a maximum depth of 2.7\%.}
    \label{fig:primary_transit}
\end{figure}

\begin{figure*}%[htpb]
    \centering
    \includegraphics[width=\linewidth]{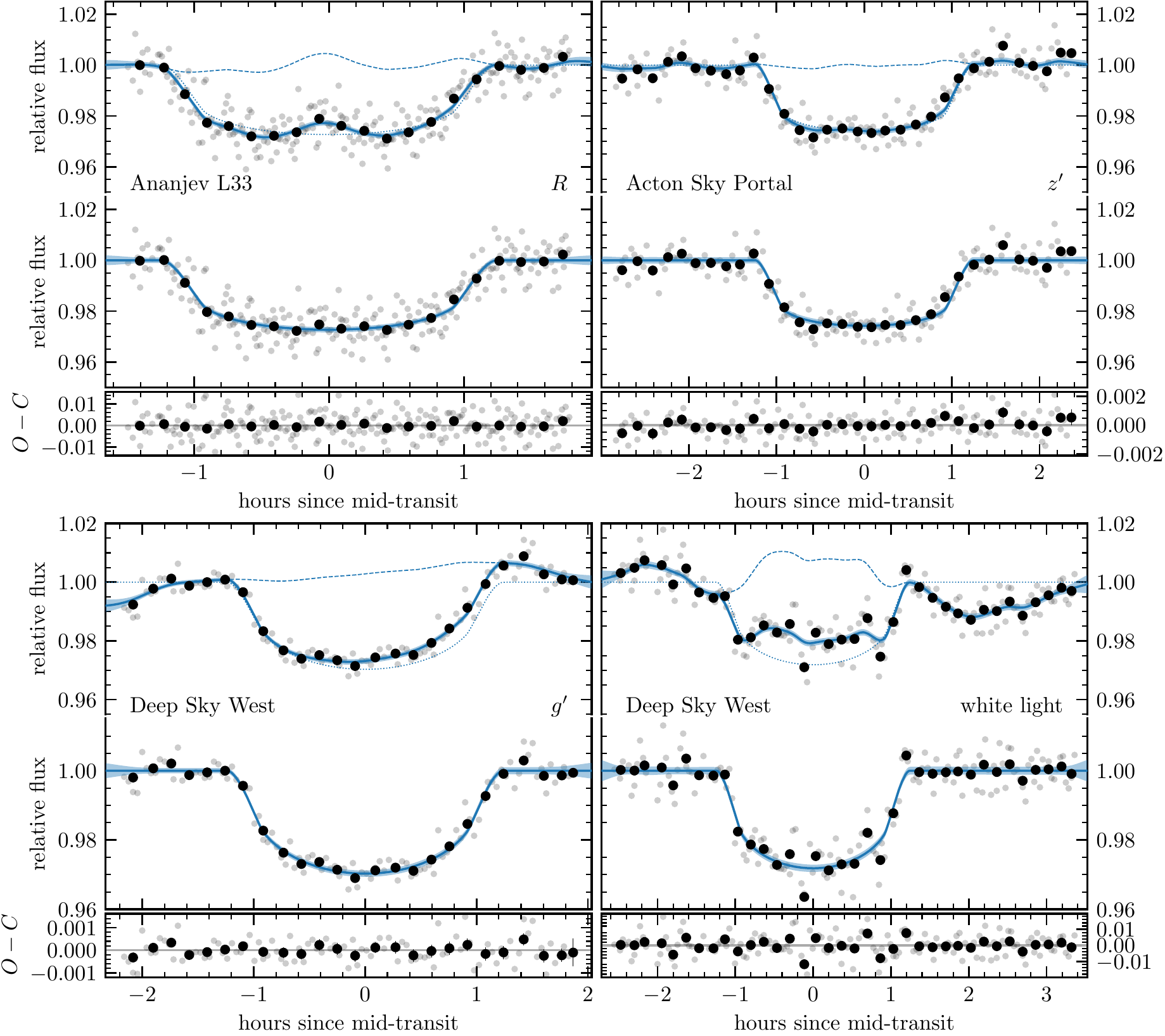}
    \caption{Transit light curves of \toip{} followed up with small ($<50\,$cm) ground-based telescopes with various filters. For each light curve, we show the raw data in the top panel (\textit{grey}) overlaid with data binned in 10 min intervals (\textit{black}). The best-fitting model is shown as a solid blue line, which consists of a Gaussian process (GP) model (\textit{dashed}) and transit model (\textit{dotted}). The $1\sigma$ uncertainty of the GP model is indicated by the shaded region. The middle panels are ``cleaned'' light curves after removing the GP model, and the bottom panels are residuals from the best-fitting model.}
    \label{fig:transits_ground}
\end{figure*}

\begin{figure*}%[htpb]
    \centering
    % \includegraphics{figures/phase_plot.pdf}
    % \caption{\emph{Top:} The light curve folded on the phase of the planet orbit, with data points binned in \SI{20}{\minute} intervals. The grey highlighted area is the transit duration. \emph{Bottom left:} Zoomed in portion of the phase folded light curve, centered on $\phi = 0.5$, where one could expect a secondary eclipse for a circular orbit. \emph{Bottom right:} A zoomed in view centered on $\phi = 0.64$, which may show a tentative signal of a secondary eclipse, at a plausible depth given the expected amount of reflected light from the planet. The location of the ``secondary eclipse" can constrain $e$ and $\omega$, which is shown in grey in Fig.~\ref{fig:ecc_weighted}. The grey rectangles denote the transit width for reference.}
    \includegraphics{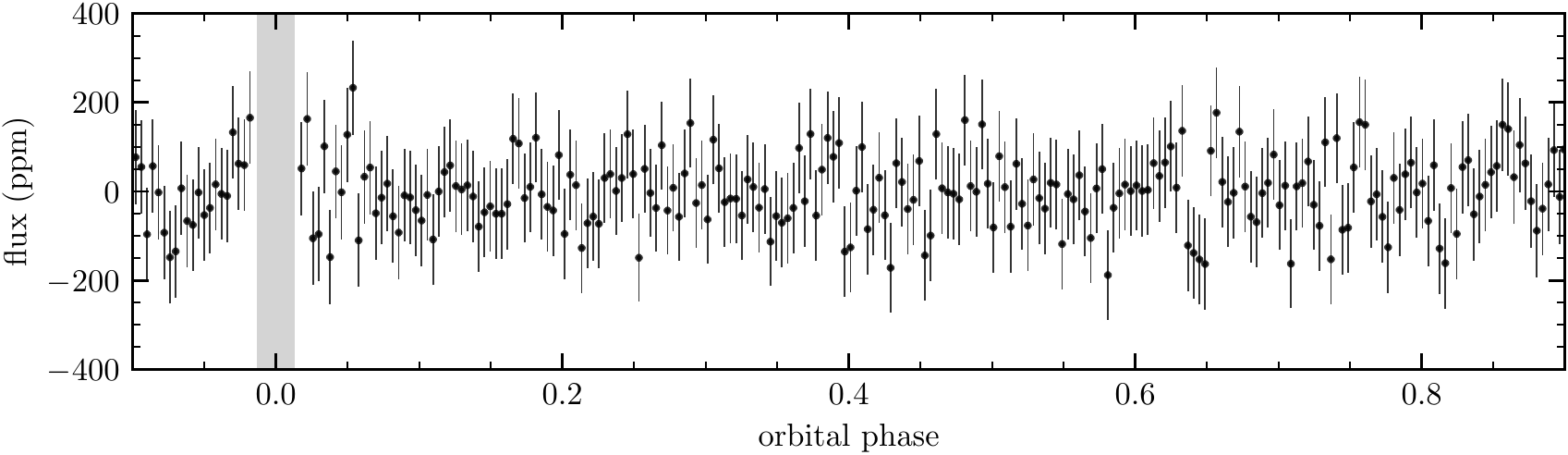}
    \caption{\emph{Top:} The light curve folded on the phase of the planet orbit, with data points binned in \SI{20}{\minute} intervals. The grey highlighted area is the transit duration. The expected secondary eclipse depth is \SI{63}{\ppm}, which is comparable to the noise. There is no apparent signal at $\phi = 0.5$, where one could expect a secondary eclipse for a circular orbit. There is a tentative signal centered on $\phi = 0.64$ with a duration that is consistent with a secondary eclipse, however this implies an eccentricity of roughly \SIrange{0.2}{0.3}{} which is incompatible with the  RV data.}
    
    \label{fig:phase_plot}
\end{figure*}

We model the combined light curves (\tess{} and ground-based) and  RV data using \exoplanet{} \citep{foreman-mackey2020}. The \exoplanet{} software uses \starry{} \citep{luger2019, agol2020} to rapidly compute analytical limb darkened light curves, and is also integrated with \celerite{} for scalable Gaussian process computations. Since the models (and their gradients) within \exoplanet{} are analytical, the software is built on the \textsc{Theano} \citep{theano} engine and therefore allows the use of \pymcthree{} \citep{salvatier2016}, which offers fast and effective convergence using gradient-based sampling algorithms.
% The advantage of using \exoplanet{} becomes clear for very complex models, such as in our case. Our full model consists of a Gaussian process model for the correlated noise and stellar variability, a transit and secondary eclipse model for the planet, and a phase curve model for the planet atmosphere that varies throughout its orbit due to being heated by the star.

The Simple Aperture Photometry (SAP) light curve from \tess{} shows a clear rotational signal of the host star (Fig.~\ref{fig:sip}). While we could take advantage of the rotational signal in our transit modelling, we derive a rotation period in Section~\ref{subsec:results_host_rotation} using the Systematics-Insensitive Periodogram (SIP). We therefore opt to use the PDCSAP flux \citep{Stumpe2012,smith2012,Stumpe2014,jenkins2016}. in our transit analysis, which is corrected for spacecraft systematics and the rotation signal since the derived rotation period is on a similar timescale to a \tess{} sector.

% Second, in removing the spot modulation the PDCSAP lightcurves have introduced some low frequency systematics that differ from sector to sector. For these reasons, we use only the SAP light curves and apply our own Gaussian processes to clean them.

 The \tess{} spacecraft fires its thrusters to unload angular momentum from its reaction wheels every few days,  which may cause the images obtained in the timestamps to appear disjoint. To make sure the momentum dumps do not affect our further analysis, we identify the times of these events from the Data Quality Flags in the FITS files, and exclude the data obtained within four hours on either side of the thruster events.

We model the out-of-transit variability using Gaussian processes (GPs). We use the \textsc{SHOTerm} model in \textsf{celerite} \citep{foreman-mackey2017, foreman-mackey2018}, fixing the quality factor $Q = 1/\sqrt{2}$ so that the covariance function becomes 
\begin{align*}
    k(\tau) = S_0\,\omega_0\,\exp{\left(-\frac{1}{\sqrt{2}} \omega_0 \tau\right)} \cos{\left( \frac{\omega_0\,\tau}{\sqrt{2}} - \frac{\pi}{4} \right)}.
\end{align*}
We fit for the natural logarithms of the amplitude and frequency, $S_0$ and $\omega_0$. Since each \tess{} sector may have systematics on different timescales and amplitudes, we model each sector and ground-based light curve with individual GPs, and also assign individual flux scaling terms and white noise terms.

\sisetup{separate-uncertainty, multi-part-units=single}
% Now that we have determined the eccentricity is non-negligible, we proceed with our combined transit and radial velocity analysis. Again, we model the rotation using the Gaussian process rotation model. We place Gaussian priors on the stellar mass and radius the values obtained from the SED analysis in Section~\ref{sec:sed}, $M_\star = \SI{0.68 \pm 0.05}{\Msun}$, $R_\star = \SI{0.739 \pm 0.031}{\Rsun}$. As before, we fit for $P$, $T_0$ and $b$, but now the planet radius, $R_\mathrm{pl}$, and eccentricity parameters $e\cos{\omega}$ and $e\sin{\omega}$. For the radial velocity model, we directly fit for the planet mass, $M_\mathrm{pl}$; radial velocity offset $\gamma$, and a white noise term $\sigma$.
We fit the \tess{} and ground-based photometry using our Gaussian process model combined with a transit model, as well as a Keplerian model for the RV data. 
We place Gaussian priors on the stellar mass and radius using values from the SED analysis in Section~\ref{subsec:sed}, $M_\star = \SI{0.68 \pm 0.08}{\Msun}$, $R_\star = \SI{0.739 \pm 0.031}{\Rsun}$.
Further, we vary the impact parameter $b$, as well as the natural logarithms of the period $P$, mid-transit time $T_0$, planet radius $R_\mathrm{pl}$, and planet mass $M_\mathrm{pl}$. The limb darkening of the star is described by a quadratic formula, with coefficients and  uncertainties within the \tess{}, $R$, $z'$, $g'$, and white light bands determined using \textsc{pyldtk} \citep{parviainen2015} and \textsc{exofast} online tool\footnote{\url{http://astroutils.astronomy.ohio-state.edu/exofast/limbdark.shtml}\citep{eastman2013}}. \textsc{pyldtk} and \textsc{exofast} interpolate the \citet{husser2013} and \citet{claret2011} atmospheric models, respectively, where we used stellar parameters from Table~\ref{tab:summary}. 
% The computed limb darkening coefficients are
% $c_1 = 0.5349 \pm 0.0037$, $c_2 = 0.1095 \pm 0.0088$. 
We vary the limb darkening coefficients with a Gaussian prior centered on the computed values, with standard deviation
of $0.01$ and $0.02$ for $c_1$ and $c_2$, respectively. These uncertainties roughly correspond to 
twice the computed error, which we inflated to account for uncertainties in the stellar atmospheric models.
% $c_1 = 0.49 \pm 0.12$, $c_2 = 0.0032 \pm 0.0074$,
The SOPHIE  RV data is further described by the semi-amplitude $K$, eccentricity parameters $\sqrt{e}\cos{\omega}$ and $\sqrt{e}\sin{\omega}$, and additional nuisance parameters that model the offset and a white noise term that is added in quadrature to the SOPHIE uncertainties.

\begin{figure}%[htpb]
    \centering
    \includegraphics[width=\linewidth]{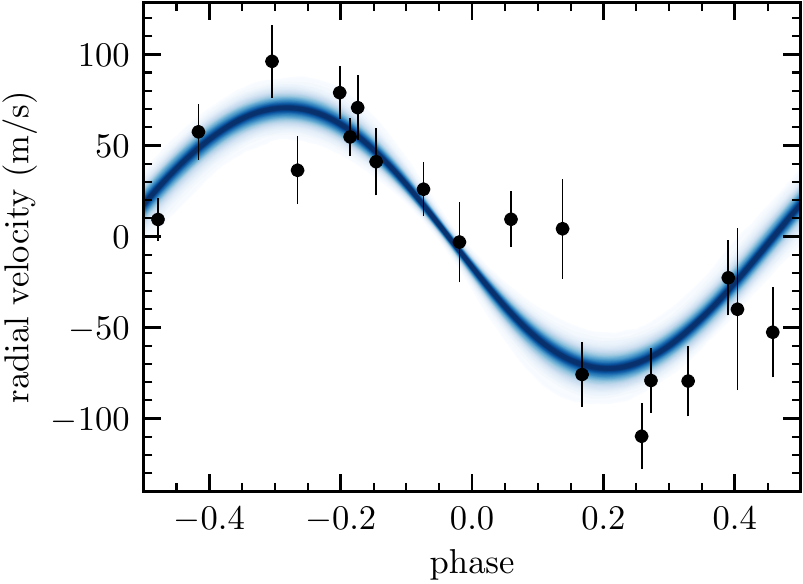}
    \caption{SOPHIE  RV (\emph{black}) with the 99\textsuperscript{th} percentile models from a joint fit with the TESS photometry encapsulated in the blue region.
    % random samples from the posterior distribution of a joint fit with the TESS photometry, allowing an eccentric orbit.
    The  RV are indicative of a $0.44M_{\rm Jup}$ planet, and are in phase with the transits observed by TESS.}
    \label{fig:radial_velocities}
\end{figure}

We first perform a maximum likelihood fit, 
% and remove points that are more than $5\sigma$ discrepant from the best-fitting model, 
followed by MCMC sampling using the NUTS sampler within \pymcthree{} to obtain credible intervals on our parameters. We launch two independent chains that are run for 4000 tuning steps and 2000 production steps. We confirmed that the sampler converged by checking the Gelman-Rubin criterion, $\hat{R} < 1.01$, and all parameters have $\num{>200}$ effective samples. 
We report the values and $15.8$ and $84.2$ percentiles for the stellar, planet and orbital parameters in Table~\ref{table:results}, and similarly for the nuisance parameters in Table~\ref{table:results_gp}. The fits to the transit photometry is shown in Fig.~\ref{fig:tess_fit},\ref{fig:primary_transit},\ref{fig:transits_ground}, and the  RV fit in Fig.~\ref{fig:radial_velocities}. 
Our  RV measurements are all published online.

% , as shown in the lower right panel in Fig.~\ref{fig:phase_plot}.
% The eccentricity distribution implied by a secondary eclipse at $\phi = 0.64 \pm 0.02$ is shown in grey in Fig.~\ref{fig:ecc_weighted}, and is in tension at $2\sigma$ with the constraints from the radial velocity data.
% We attempted to include a model for the secondary eclipse in our analysis, but were unable to constrain the location or depth of the eclipse without very strict priors.

\begin{table*}%[htbp]
\renewcommand{\arraystretch}{1.2}
%    \sisetup{round-mode=places}
    \centering
    \caption{Derived parameters from the joint modelling of \tess{} photometric and SOPHIE  RV data. $^a$Controlled by Gaussian prior, $\mathcal{N}(0.68, 0.05)$. $^b$Controlled by Gaussian prior, $\mathcal{N}(0.739, 0.031)$. 
    $^c$Assuming zero albedo.}
    \begin{tabular}{@{\extracolsep{\fill}}
    % \begin{tabular*}{\linewidth}{@{\extracolsep{\fill}}
            llcc}
        \toprule
        \toprule
        {\textbf{Parameter}} & {\textbf{Description}} & {\textbf{Value}} & {\textbf{Value}} \\% & {\textbf{Reference/data}}  \\
        & & \textit{circular model \textbf{(adopted)}} & \textit{eccentric model} \\
        \midrule
        \multicolumn{4}{@{}l@{}}{\emph{Derived stellar parameters}} \\
        \midrule
        % $T_\mathrm{eff}$ (\si{\kelvin}) & Effective temperature & $6037 \pm 45$ & \text{\cite{ghezzi2010}}\\
        % $\log{g}$ (cgs) & Stellar surface gravity & $4.42 \pm 0.03$ & \text{\cite{ghezzi2010}}\\
        % $[\mathrm{Fe/H}]$ (dex) & Stellar metallicity & $0.08 \pm 0.03$ & \text{\cite{ghezzi2010}}\\
        % \vsini{} (\si{\kilo\metre\per\second}) & Projected rotational velocity & $3.14 \pm 0.50^{\textblue{a}}$ &
        % \text{\cite{valenti2005}} \\
        % $T_\mathrm{eff}$ (\si{\kelvin}) & Effective temperature &  $4699 \pm 130$ & \\
        % $[\mathrm{Fe/H}]$ (dex) & Stellar metallicity & $0.09 \pm 0.04$ & \text{\cite{damasso2020}}\\
        % \vsini{} (\si{\kilo\metre\per\second}) & Projected rotational velocity & $3.34 \pm 0.07^{\textblue{a}}$ &
        % \text{\cite{damasso2020}} \\
        $M_\star$ (\si{\Msun}) & Stellar mass$^a$ & $0.744^{+0.064}_{-0.059}$ & $0.743^{+0.066}_{-0.064}$ \\
        $R_\star$ (\si{\Rsun}) & Stellar radius$^b$ & $0.711^{+0.020}_{-0.019}$ & $0.711^{+0.024}_{-0.024}$ \\
        $\rho_\star$ (\si{\gram\per\cubic\centi\metre}) & Stellar density & $2.92^{+0.03}_{-0.04}$ &  $2.91^{+0.26}_{-0.25}$ \\
        $\log{g_\star}$ (cgs) & Stellar surface gravity & $4.605^{+0.013}_{-0.013}$ & $4.581^{+0.044}_{-0.045}$ \\[5pt]
        % $P_\text{rot}$ (days) & Photometric rotation period$^c$ & $32.4^{+5.9}_{-3.7}$ &  $41.5^{+5.7}_{-7.1}$ \\[5pt]

        \multicolumn{3}{@{}l@{}}{\emph{Derived planet parameters}} \\
        \midrule

        $P$ (days)  & Orbital period & $3.4779780^{+0.0000019}_{-0.0000017}$ &  $3.4779779^{+0.0000018}_{-0.0000016}$ \\
        
        $T_0$ (BJD$_\mathrm{UTC}$ - \num{2457000}) & Transit mid-point & $1686.700531^{+0.000097}_{-0.000104}$ &
        $1686.700536^{+0.000099}_{-0.000100}$ \\
        $M_\text{pl}$ (\si{\Mjup}) & Planet mass & $0.441^{+0.049}_{-0.047}$ & $0.440^{+0.051}_{-0.047}$ \\
        $R_\text{pl}$ (\si{\Rjup}) & Planet radius & $1.022^{+0.030}_{-0.027}$ & $1.021^{+0.034}_{-0.034}$  \\
        $\rho_\text{pl}$ (\si{\gram\per\cubic\centi\metre}) & Planet density & $0.513^{+0.051}_{-0.048}$ & $0.513^{+0.071}_{-0.066}$  \\
        $\log{g_\text{pl}}$ (cgs) & Planet surface gravity & $3.019^{+0.040}_{-0.040}$ & $3.019^{+0.050}_{-0.053}$ \\
        $M_\text{pl}/M_\star$ & Mass ratio & $0.000567^{+0.000056}_{-0.000055}$ &  $0.000568^{+0.000052}_{-0.000051}$  \\
        $R_\text{pl}/R_\star$ & Planet-to-star radius ratio & $0.14762^{+0.00035}_{-0.00030}$ &
        $0.14764^{+0.00032}_{-0.00031}$  \\
        $a/R_\star$ & Scaled separation & $12.314^{+0.036}_{-0.056}$ & $12.301^{+0.352}_{-0.358}$  \\
        $R_\star/a$ & Scaled stellar radius & $0.08121^{+0.00037}_{-0.00023}$ &  $0.08130^{+0.00244}_{-0.00226}$  \\
        $R_\text{pl}/a$ & Scaled planet radius & $0.011981^{+0.000077}_{-0.000033}$ &  $0.012002^{+0.000371}_{-0.000348}$  \\
        $T_\mathrm{eq}$ (\si{\kelvin}) & Planet equilibrium temperature$^c$ & $963^{+21}_{-21}$ &  $963^{+25}_{-25}$  \\
        $b$ ($R_\star$) & Impact parameter & $0.065^{+0.055}_{-0.044}$ &  $0.064^{+0.055}_{-0.045}$ \\
        $i_\text{p}$ (\si{\degree}) & Orbital inclination &$89.70^{+0.20}_{-0.26}$ & $89.70^{+0.21}_{-0.26}$  \\
        $a$ (AU) & Semi-major axis & $0.04070^{+0.00114}_{-0.00110}$ &  $0.04069^{+0.00116}_{-0.00120}$ \\
        $D_{T_0}$ & Transit depth at $T_0$ & $0.026759^{+0.000099}_{-0.000100}$ &  $0.026756^{+0.000097}_{-0.000095}$ \\
        $T_{14}$ (days) & Transit duration between $1\textsuperscript{st}$ and $4\textsuperscript{th}$ contacts & $0.10314^{+0.00020}_{-0.00020}$ &  $0.10313^{+0.00020}_{-0.00019}$  \\
        $K$ (\si{\metre\per\second}) & RV semi-amplitude & $72.0^{+6.8}_{-6.4}$ & $72.2^{+6.4}_{-6.4}$ \\
        $e$ & Eccentricity & $0$ &  $0.030^{+0.034}_{-0.022}$ \\
        $\omega$ (\si{\degree}) & Argument of periastron & -- & $5^{+132}_{-138}$  \\
        $\sqrt{e}\cos{\omega}$ & Unit eccentricity parameter & -- & $-0.038^{+0.142}_{-0.127}$ \\
        $\sqrt{e}\sin{\omega}$ & Unit eccentricity parameter & -- & $0.005^{+0.144}_{-0.156}$ \\
        $u_\mathrm{TESS} $ & Limb darkening coefficient, \tess{} band & $0.5241^{+0.0066}_{-0.0064}$ & $0.5232^{+0.0067}_{-0.0068}$ \\
        $v_\mathrm{TESS} $ & Limb darkening coefficient, \tess{} band & $0.088^{+0.016}_{-0.017}$ & $0.088^{+0.015}_{-0.015}$ \\ %[5pt]
        $u_R $ &Limb darkening coefficient, $R$ band & $0.5328^{+0.0103}_{-0.0097}$ & $0.5333^{+0.0099}_{-0.0098}$ \\
        $v_R$ & Limb darkening coefficient, $R$ band & $0.171^{+0.020}_{-0.020}$ & $0.172^{+0.020}_{-0.020}$ \\
        $u_{z'} $ & Limb darkening coefficient, $z'$ band  & $0.3689^{+0.0102}_{-0.0097}$ & $0.3687^{+0.0100}_{-0.0096}$ \\
        $v_{z'}$ & Limb darkening coefficient, $z'$ band & $0.209^{+0.020}_{-0.020}$ &$0.209^{+0.021}_{-0.020}$ \\
        $u_{g'} $ & Limb darkening coefficient, $g'$ band & $0.8026^{+0.0098}_{-0.0101}$  & $0.8024^{+0.0100}_{-0.0098}$ \\
        $v_{g'}$ & Limb darkening coefficient, $g'$ band & $0.030^{+0.020}_{-0.021}$ & $0.029^{+0.021}_{-0.020}$ \\
        $u_\mathrm{white}$ & Limb darkening coefficient, white light & $0.6640^{+0.0098}_{-0.0100}$ & $0.6627^{+0.0105}_{-0.0104}$ \\
        $v_\mathrm{white}$ & Limb darkening coefficient, white light & $0.059^{+0.019}_{-0.019}$ & $0.062^{+0.021}_{-0.020}$ \\
        \bottomrule
    % \end{tabular*}
    \end{tabular}
    \label{table:results}
\end{table*}

The combined fit to the \tess{} light curves and SOPHIE RVs reveals that \toip{} is a giant planet with mass $M_\mathrm{pl} = 0.441^{+0.049}_{-0.047}\,\si{\Mjup}$, and radius $R_\text{pl} =  1.022^{+0.030}_{-0.027}\,\si{\Rjup}$. The  RV data constrains the eccentricity to be ${<}0.13$ ($99\%$ credible interval). Using the approximate tidal circularization timescale in §6 of \citet{barker2009} with their example tidal parameters, we find that \toip{} might have circularized after ${\sim}2\,$Gyr, which is shorter than the estimated age of the host star (${\sim}5\,$Gyr). As in Section~\ref{subsec:results_rvfit}, the eccentricity measurement from our global analysis is not statistically signficant. The BIC prefers a circular model over the eccentric model, with $\Delta\mathrm{BIC} = \mathrm{BIC}_\mathrm{circ} - \mathrm{BIC}_\mathrm{ecc} = -23.8$. We therefore adopt the circular model, but report results from both analyses in Table~\ref{table:results}.

We visually searched for a secondary eclipse in the binned residuals of the phase-folded TESS photometry, shown in Fig.~\ref{fig:phase_plot}. The residuals close to phase 0.5 (where a secondary eclipse would be for a circular orbit) show no signs of a planet occultation. There may be a tentative signal of a \SI{\sim 100}{\ppm} secondary eclipse at phase $\sim0.64$. However, this implies an eccentricity of roughly \SIrange{0.2}{0.3}{}. Such high eccentricities are not supported by the current  RV data. 
We estimate the secondary eclipse depth to be 63 ppm, using the simple approximation $\approx0.5(R_{\rm pl}/a_{\rm pl})^2$, where we assume a geometric albedo of 0.5 and ignore thermal emission. An eclipse of this depth would be consistent with the small dip seen at phase 0.64, but more observations, potentially including those from an extended TESS mission, would be needed to confirm that this feature is a secondary eclipse.

\section{Discussion}\label{sec:discussion}

 We have confirmed that the {\it TESS} transiting candidate TOI-1259Ab is a 0.441$M_{\rm Jup}$ transiting exoplanet on a 3.48 day orbit, through our  RV and ground-based photometric follow-up. Furthermore, by combining with the existing {\it Gaia} binaries catalog of \citet{elbadry2018}, we show that this planet exists in a binary star system, where its primary star is a $0.68M_{\odot}$ K-dwarf and the secondary star is a $0.56M_{\odot}$ white dwarf on a bound orbit. The current projected separation is roughly 1648 AU. All of the key parameters of this system are summarised in Table~\ref{tab:summary}.

\subsection{Comparison with \citet{mugrauer2020}}\label{subsec:discussion_mugrauer}

\citet{mugrauer2020} independently characterised the white dwarf companion to TOI-1259. They also concluded that it was a bound companion, and determined a projected separation of roughly 1600 AU, which is the same as in \citet{elbadry2018}. Their derived white dwarf effective temperature of $T_{\rm eff} = 6473^{+672}_{-419}$\,K agrees with our calculations of $T_{\rm eff} = 6300^{+80}_{-70}$\,K.

\subsection{Dynamical history}\label{subsec:discussion_dynamics}

\citet{wang2014} determined that the planet frequency in binary systems was lower than that around single stars for binary separations up to 1500 AU. Some other studies suggest that the influence of a binary is less far-reaching, with only ~100-200 AU and tighter binaries affecting planet populations \citep{kraus2016,moe2019,ziegler2021}. With a mass of $0.56M_{\odot}$ at a distance of about 1600 AU, our white dwarf is presently at a separation not predicted to impact planet formation. However, during its main sequence lifetime, the white dwarf's progenitor would have been both more massive ($\sim1.59M_{\odot}$) and much closer ($\sim900$ AU, assuming adiabatic mass loss). At this point secular effects such as Kozai-Lidov \citep{lidov1962,kozai1962,mazeh1979} may have been relevant to the planet. Indeed, the Kozai-Lidov effect may have brought the planet to its current orbital configuration, by inducing high-eccentricity tidal migration \citep{fabrycky2007,naoz2012,naoz2016}, if the planet started out at a wider orbit, as would be expected for a gas giant. Given the estimated pre-white dwarf age of the system ($~2$ Gyr), a large range of orbital parameters could have led to the observed orbit of the planet TOI-1259Ab. In such a case, the companion star evolving into a white dwarf may have also acted as a natural ``shutoff'' for such secular effects \citep{Dawson2018}.  

 We also note here that our combined photometry and RV fit favours a circular solution for the planetary orbit. The planet may have had a higher eccentricity (e.g. due to Kozai-Lidov) in the past but with a semi-major axis of only $0.04$ AU it likely would have been circularized by tidal interactions within the age of the system. Higher precision photometric follow-up that is able to reveal the secondary transit of the planet would most likely be the best means of detecting any potential small but non-zero eccentricity.

Any planets that orbited the white dwarf progenitor may have experienced the opposite effect; the evolution of their host into a white dwarf may have acted to ``turn on'' secular effects \citep{shappee2013,stephan2017,stephan2018,stephan2020b}. This could cause the destruction of its planets. TOI-1259A's white dwarf companion would therefore be a worthwhile target for finding signatures of heavy element pollution by planetary debris. In any case, more detailed studies of the dynamical history of TOI-1259 and of Gliese-86b, which has a white dwarf companion at a projected separation of just 21 AU \citep{queloz2000,els2001,lagrange2006}, are warranted and will be part of a future work.

\subsection{Future spectroscopic characterisation of the white dwarf companion}\label{subsec:discussion_WDspectrum}

There is a propensity for white dwarfs to have atmospheres contaminated by heavy elements \citep{debes2012,farihi2016,wilson2019}, despite an expectation that such elements would quickly settle towards the core due to the high gravity. This has been measured in white dwarfs without any known exoplanet companions, but the pollution itself has been attributed to the accretion of surrounding planetary material. It would be interesting to know if WD such as TOI-1259B, which do have a known associated planet, are more likely to be polluted than ``lonely'' WDs. \citet{southworth2020} most recently tested this for WASP-98\footnote{A similar configuration to TOI-1259, but with a wider binary separated by $\sim3500$ AU.}, but their spectroscopy of the white dwarf revealed a featureless spectrum and hence no evidence of pollution could be ascertained. It will ultimately be beneficial to conduct such spectroscopy on not only TOI-1259B, but indeed all of the similar systems listed in Table~\ref{tab:known_planets}.

Follow-up spectroscopy will also hopefully inform us if the atmosphere is hydrogen- or helium-dominated. Breaking this degeneracy would allow a more accurate constraint on the WD parameters, since we could choose the appropriate model from Table~\ref{tab:wd_fit}.

\subsection{Future JWST observations}\label{subsec:discussion_jwst}

\citet{kempton2018} derived a Transmission Spectroscopy Metric (TSM). It is a means of prioritizing exoplanets for future atmospheric characterisation, in particular using the James Webb Space Telescope, and is proportional to the expected signal to noise ratio for the planet's transmission spectrum. Their analytic expression is

\begin{equation}
    \label{eq:tsm}
    {\rm TSM} = S \times \frac{R_{\rm pl}^3T_{\rm eq}}{M_{\rm pl}R_{\star}^2}\times10^{-m_{\rm J}/5},
\end{equation}
where $m_{\rm J}$ is the magnitude of the host star in the J band and $S$ is a an empirical scale factor derived by \citet{kempton2018} to make their simple analytic expression match the more detailed simulations of \citet{louie2020}. The factor $S$ depends on the radius of the transiting planet: $S=0.19$ for $R_{\rm pl}<1.5R_{\oplus}$; $S=1.26$ for $1.5<R_{\rm pl} < 2.75R_{\oplus}$; $S=1.28$ for $2.75<R_{\rm lp} < 4.0R_{\oplus}$ and $S=1.15$ for $R_{\rm pl} > 4.0R_{\oplus}$\footnote{Whilst we use this final scale factor for all planets above $4R_{\oplus}$, \citet{kempton2018} only define it for 4 to 10$R_{\oplus}$ and do not give a scale factor for larger planets.}.

\begin{figure}%[htpb]
    \centering
    \includegraphics[width=\linewidth]{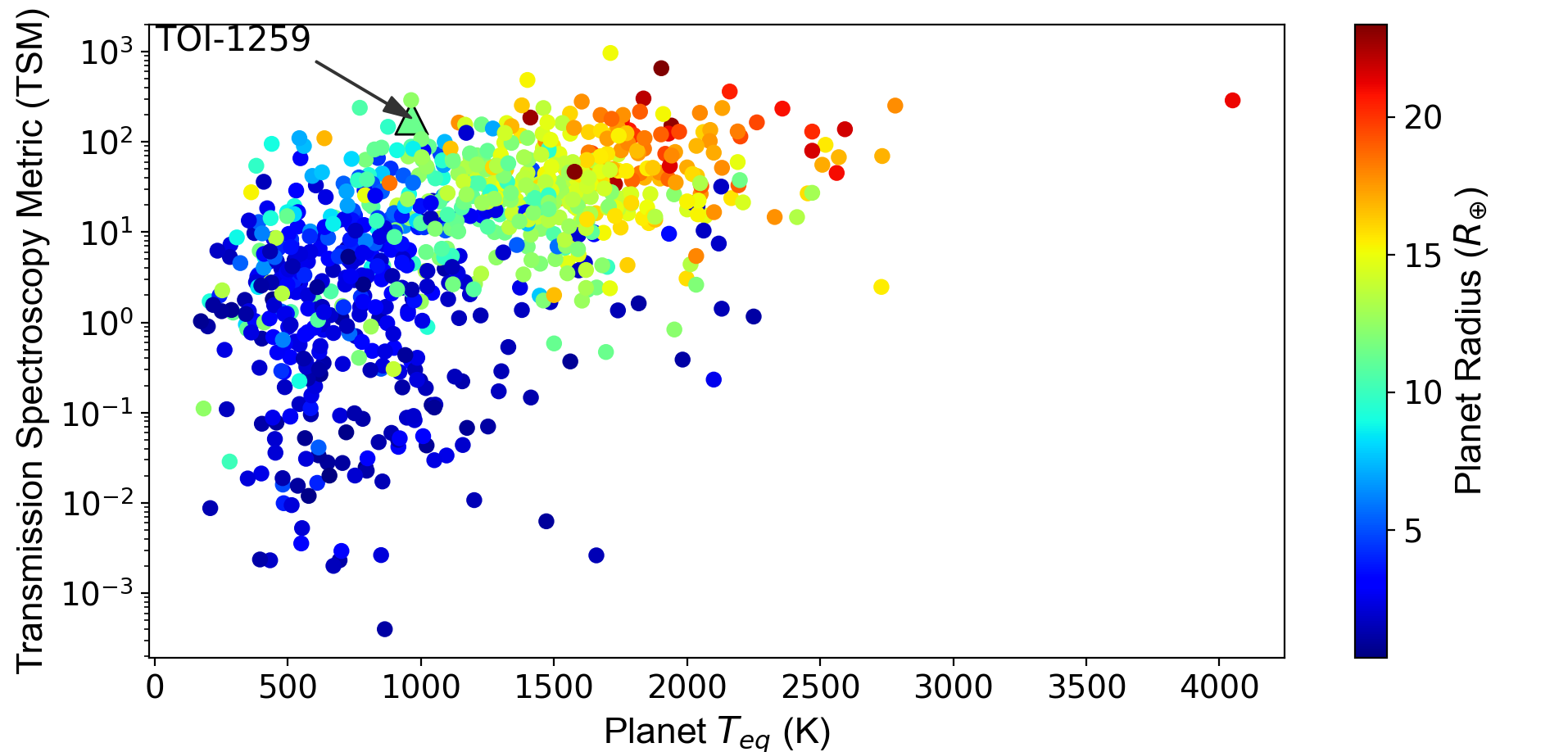}
    \caption{Transit Spectroscopy Metric (TSM) from \citet{kempton2018} for all confirmed transiting exoplanets as a function of their equilibrium temperature (where such a value has been calculated). The TSM is calculated using Eq.~\ref{eq:tsm}. The color scale is the planet radius. TOI-1259Ab has a TSM of 180, making it ideal for atmospheric follow-up. In particular, it has one of the highest TSM values for a planet cooler than 1000 K and of Jupiter-size or less.}
    \label{fig:tsm_scatter}
\end{figure}

For TOI-1259Ab we calculate a value of TSM $=180$, which places it in the top 3\% of all confirmed transiting exoplanets. We demonstrate this in Fig.~\ref{fig:tsm_scatter}. TOI-1259Ab has a scaled separation $a/R_{\star}=12.314^{+0.036}_{-0.056}$ and an equilibrium temperature $T_{\rm eq}=963\pm21$ K. Most of the planets with a higher TSM value are hotter and larger. There are only two objects cooler than 1000 K with a higher TSM:  WASP-69b and WASP-107b. These two planets have already received considerable attention with respect to their atmospheres.  WASP-69b \citep{Anderson2014} has a confirmed presence of helium in its atmosphere from ground-based observations \citep{Nortmann2018}. WASP-107b \citep{anderson2017} is a so-called ``super-puff'', based on a $0.12M_{\rm Jup}$ mass and a $0.94R_{\rm Jup}$ radius, and Hubble Space Telescope observations have revealed the presence of both helium \citep{spake2018} and water \citep{kriedberg2018}.

With an ecliptic latitude of $76.878^{\circ}$, TOI-1259Ab is near the JWST continuous viewing zone and is observable for typically 227 days per year, which will assist future  atmospheric characterisation.

\section*{Affiliations}

% List of institutions
$^{1}$Department of Astronomy, The Ohio State University, Columbus, OH 43210, USA\\
$^{2}$Department of Astronomy and Theoretical Astrophysics Center, University of California Berkley, Berkley, CA 94720, USA\\
$^{3}$School of Physics and Astronomy, University of Birmingham, Edgbaston, Birmingham B15 2TT, UK\\
$^{4}$American Museum of National History, New York, NY 10024, USA \\
$^{5}$Center for Computational Astrophysics, Flatiron Institute, New York, NY 10010, USA \\
$^{6}$Department of Astronomy, University of Washington, Seattle, WA 98105, USA \\
$^{7}$Bay Area Environmental Research Institute, P.O. Box 25, Moffett Field, CA, USA\\
$^{8}$School of Physics, University of New South Wales, Sydney, NSW 2052, Australia\\
$^{9}$Sydney Institute for Astronomy (SIfA), School of Physics, University of Sydney, NSW 2006, Australia\\
$^{10}$Aix Marseille Univ, CNRS, CNES, LAM, Marseille, France\\
$^{11}$Department of Physics \& Astronomy, Vanderbilt University, Nashville, TN 37235, USA\\
$^{12}$Center for Cosmology and AstroParticle Physics, The Ohio State University, Columbus, OH 43210, USA\\ 
$^{13}$Acton Sky Portal Private Observatory, Acton, MA, USA\\
$^{14}$Laboratory of Astrochemical Research, Ural Federal University, Ekaterinburg, Russia, ul. Mira d. 19, Yekaterinburg, Russia, 620002\\
$^{15}$Astronomical department, Ural Federal University, Yekaterinburg, Russia\\
$^{16}$Private Astronomical Observatory, Ananjev, Odessa Region, 66400, Ukraine\\
$^{17}$Department of Physics, Engineering and Astronomy,  Stephen F. Austin State University, TX 75962, USA\\
$^{18}$Oukaimeden Observatory, High Energy Physics and Astrophysics Laboratory, Cadi Ayyad University, Marrakech, Morocco\\
$^{19}$NASA Ames Research Center, MS 244-30, Moffett Field, CA 94035, USA\\
$^{20}$SETI Institute, 189 Bernardo Ave, Suite 200, Mountain View, CA 94043, USA\\
$^{21}$Harvard-Smithsonian Center for Astrophysics, 60 Garden St, Cambridge, MA, 02138, USA\\
$^{22}$Department of Physics and Kavli Institute for Astrophysics and Space Research, Massachusetts Institute of Technology, Cambridge, MA 02139, USA\\
$^{23}$Proto-Logic LLC, 1718 Euclid Street NW, Washington, DC 20009, USA\\
$^{24}$Space Telescope Science Institute, 3700 San Martin Drive, Baltimore, MD, 21218, USA\\
$^{25}$Department of Earth, Atmospheric and Planetary Sciences, Massachusetts Institute of Technology, Cambridge, MA 02139, USA\\
$^{26}$Department of Aeronautics and Astronautics, Massachusetts Institute of Technology, Cambridge, MA 02139, USA\\
$^{27}$Department of Astronomy, The University of Wisconsin-Madison, Madison, WI 53706, USA\\
$^{28}$Department of Astrophysical Sciences, 4 Ivy Lane, Princeton University, Princeton, NJ 08544, USA\\
$^{29}$Astrophysics Group, Keele University, Keele, Staffordshire, ST5 5BG, UK
$^{30}$Centre for Exoplanets and Habitability, University of Warwick, Gibbet Hill Road, Coventry CV4 7AL, UK
% $\dagger$ Fulbright Fellow \\
% $\ddagger$ Fellow of the Swiss National Science Foundation

\section*{Acknowledgements}
This project began at the Expanding the Science of TESS meeting, which took place in February 2020 the University of Sydney, back when meeting people in large groups was still a thing. The  RV observations were partly conducted while OHP was in ``remote observing'' mode, a special mode produced as a response the unique COVID-19 situation. We are extremely grateful for the dedication of the staff at OHP that allowed observations to resume. We thank Markus Mugrauer for looking at a draft version of this paper. Finally, we thank a referee for providing a thorough review which undoubtedly improved the quality of the paper.

The observations were obtained under an OHP DDT programme (PI Triaud). This work was in part funded by the U.S.--Norway Fulbright Foundation and a NASA \tess{} GI grant G022253 (PI: Martin). DVM received funding from the Swiss National Science Foundation (grant number  P 400P2 186735). AHMJT  received funding from the European Research Council (ERC) under the European Union’s Horizon 2020 research and innovation programme (grant agreement n$^\circ$ 803193/BEBOP). VKH is also supported by a Birmingham Doctoral Scholarship, and by a studentship from Birmingham's School of Physics \& Astronomy. SG has been supported by STFC through consolidated grants ST/L000733/1 and ST/P000495/1. SJM was supported by the Australian Research Council through DECRA DE180101104. VK  has been supported by the Ministry of science and higher education of Russian Federation, topic № FEUZ-2020-0038.

We acknowledge the use of public TESS Alert data from pipelines at the TESS Science Office and at the TESS Science Processing Operations Center. Some of the data presented in this paper were obtained from the Mikulski Archive for Space Telescopes (MAST).
Resources supporting this work were provided by the NASA High-End Computing (HEC) Program through the NASA Advanced Supercomputing (NAS) Division at Ames Research Center for the production of the SPOC data products.

This research made use of \textsf{AstroPy},\footnote{http://www.astropy.org} a community-developed core Python package for Astronomy \citep{astropy2013, astropy2018}.

\section*{Data availability}
All TESS data are publicaly available and can be downloaded using \lightkurve{} \citep{lightcurve2018} or other tools. The SOPHIE RVs are published as supplementary data. Any other data/models in this article will be shared on reasonable request to the corresponding author.

\bibliographystyle{mnras}
\bibliography{references}

%%%%%%%%%%%%%%%%% APPENDICES %%%%%%%%%%%%%%%%%%%%%
% \newpage
\appendix

\section{Nuisance parameters from the global analysis}

\begin{table*}%[htbp]
\renewcommand{\arraystretch}{1.2}
%    \sisetup{round-mode=places}
    \centering
    \caption{Fitted Gaussian process (GP) and radial velocity nuisance parameters from the joint modelling of the photometric and radial velocity data. Only the parameters for the circular orbital model are given, as the differences from the eccentric model are negligible.}
    \begin{tabular}{@{\extracolsep{\fill}}
    % \begin{tabular*}{\linewidth}{@{\extracolsep{\fill}}
            llcc}
        \toprule
        \toprule
        {\textbf{Parameter}} & {\textbf{Description}} & {\textbf{Value}} &  {\textbf{Prior}}\\% & {\textbf{Reference/data}}  \\
        \midrule
        \multicolumn{4}{@{}l@{}}{\emph{Sector 14}} \\
        % \midrule
        $\log{S_0\omega_0^4}$ & GP amplitude & $-15.77^{+0.59}_{-0.51}$ & $\mathcal{N}(\log{\mathrm{var(flux)}}, 5^2)$ \\
        $\log{\omega_0}$ (\si{\per\second}) & GP frequency & $-0.90^{+0.56}_{-1.20}$ & $\mathcal{N}(\log{2\pi / 5}, 5^2)$ \\
        $\log{\sigma^2}$ & White noise term & $-24.52^{+3.16}_{-5.73}$ & $\mathcal{N}(\log{\mathrm{var(flux)}}, 10^2)$ \\
        $\mathcal{F}$ & Flux scaling factor & $0.99949^{+0.00053}_{-0.00446}$ & $\mathcal{N}(\mathrm{median(flux)}, 5^2)$ \\[2pt]
        \multicolumn{4}{@{}l@{}}{\emph{Sector 17}} \\
        % \midrule
        $\log{S_0\omega_0^4}$ & GP amplitude & $-8.71^{+0.28}_{-0.27}$ & $\mathcal{N}(\log{\mathrm{var(flux)}}, 5^2)$ \\
        $\log{\omega_0}$ (\si{\per\second}) & GP frequency & $0.28^{+0.32}_{-0.38}$ & $\mathcal{N}(\log{2\pi / 5}, 5^2)$ \\
        $\log{\sigma^2}$ & White noise term & $-24.52^{+3.29}_{-5.28}$ & $\mathcal{N}(\log{\mathrm{var(flux)}}, 10^2)$ \\
        $\mathcal{F}$ & Flux scaling factor & $1.00051^{+0.00206}_{-0.00247}$ & $\mathcal{N}(\mathrm{median(flux)}, 5^2)$ \\[2pt]
        \multicolumn{4}{@{}l@{}}{\emph{Sector 18}} \\
        % \midrule
        $\log{S_0\omega_0^4}$ & GP amplitude & $-10.19^{+0.33}_{-0.31}$ & $\mathcal{N}(\log{\mathrm{var(flux)}}, 5^2)$ \\
        $\log{\omega_0}$ (\si{\per\second})  & GP frequency & $0.75^{+0.16}_{-0.17}$ & $\mathcal{N}(\log{2\pi / 5}, 5^2)$ \\
        $\log{\sigma^2}$ & White noise term & $-24.86^{+3.47}_{-6.10}$ & $\mathcal{N}(\log{\mathrm{var(flux)}}, 10^2)$ \\
        $\mathcal{F}$ & Flux scaling factor & $0.99975^{+0.00041}_{-0.00044}$ & $\mathcal{N}(\mathrm{median(flux)}, 5^2)$ \\[2pt]
        \multicolumn{4}{@{}l@{}}{\emph{Sector 19}} \\
        % \midrule
        $\log{S_0\omega_0^4}$ & GP amplitude & $-14.68^{+0.50}_{-0.52}$ & $\mathcal{N}(\log{\mathrm{var(flux)}}, 5^2)$ \\
        $\log{\omega_0}$ (\si{\per\second}) & GP frequency & $-0.01^{+0.25}_{-0.29}$ & $\mathcal{N}(\log{2\pi / 5}, 5^2)$ \\
        $\log{\sigma^2}$ & White noise term & $-24.74^{+3.25}_{-5.49}$ & $\mathcal{N}(\log{\mathrm{var(flux)}}, 10^2)$ \\
        $\mathcal{F}$ & Flux scaling factor & $1.00004^{+0.00019}_{-0.00019}$ & $\mathcal{N}(\mathrm{median(flux)}, 5^2)$ \\[2pt]
        \multicolumn{4}{@{}l@{}}{\emph{Sector 20}} \\
        % \midrule
        $\log{S_0\omega_0^4}$ & GP amplitude & $-14.21^{+0.39}_{-0.37}$ & $\mathcal{N}(\log{\mathrm{var(flux)}}, 5^2)$ \\
        $\log{\omega_0}$ (\si{\per\second}) & GP frequency & $-0.37^{+0.24}_{-0.35}$ & $\mathcal{N}(\log{2\pi / 5}, 5^2)$ \\
        $\log{\sigma^2}$ & White noise term & $-24.89^{+3.19}_{-5.37}$ & $\mathcal{N}(\log{\mathrm{var(flux)}}, 10^2)$ \\
        $\mathcal{F}$ & Flux scaling factor & $0.99995^{+0.00044}_{-0.00061}$ & $\mathcal{N}(\mathrm{median(flux)}, 5^2)$ \\[2pt]
        \multicolumn{4}{@{}l@{}}{\emph{Sector 21}} \\
        % \midrule
        $\log{S_0\omega_0^4}$ & GP amplitude & $-11.36^{+0.38}_{-0.40}$ & $\mathcal{N}(\log{\mathrm{var(flux)}}, 5^2)$ \\
        $\log{\omega_0}$ (\si{\per\second}) & GP frequency & $1.04^{+0.15}_{-0.16}$ & $\mathcal{N}(\log{2\pi / 5}, 5^2)$ \\
        $\log{\sigma^2}$ & White noise term & $-24.77^{+3.13}_{-5.82}$ & $\mathcal{N}(\log{\mathrm{var(flux)}}, 10^2)$ \\
        $\mathcal{F}$ & Flux scaling factor & $1.00006^{+0.00012}_{-0.00013}$ & $\mathcal{N}(\mathrm{median(flux)}, 5^2)$ \\[2pt]
        \multicolumn{4}{@{}l@{}}{\emph{Sector 24}} \\
        % \midrule
        $\log{S_0\omega_0^4}$ & GP amplitude & $-11.28^{+0.42}_{-0.44}$ & $\mathcal{N}(\log{\mathrm{var(flux)}}, 5^2)$ \\
        $\log{\omega_0}$ (\si{\per\second}) & GP frequency & $1.14^{+0.16}_{-0.17}$ & $\mathcal{N}(\log{2\pi / 5}, 5^2)$ \\
        $\log{\sigma^2}$ & White noise term & $-24.71^{+3.26}_{-5.39}$ & $\mathcal{N}(\log{\mathrm{var(flux)}}, 10^2)$ \\
        $\mathcal{F}$ & Flux scaling factor & $1.00007^{+0.00010}_{-0.00011}$ & $\mathcal{N}(\mathrm{median(flux)}, 5^2)$ \\[2pt]
         \multicolumn{4}{@{}l@{}}{\emph{Sector 25}} \\
        % \midrule
        $\log{S_0\omega_0^4}$ & GP amplitude & $-9.83^{+0.34}_{-0.32}$ & $\mathcal{N}(\log{\mathrm{var(flux)}}, 5^2)$ \\
        $\log{\omega_0}$ (\si{\per\second}) & GP frequency & $1.51^{+0.13}_{-0.13}$ & $\mathcal{N}(\log{2\pi / 5}, 5^2)$ \\
        $\log{\sigma^2}$ & White noise term & $-24.64^{+3.34}_{-5.93}$ & $\mathcal{N}(\log{\mathrm{var(flux)}}, 10^2)$ \\
        $\mathcal{F}$ & Flux scaling factor & $1.00006^{+0.00010}_{-0.00010}$ & $\mathcal{N}(\mathrm{median(flux)}, 5^2)$ \\[2pt]
        \multicolumn{4}{@{}l@{}}{\emph{Sector 26}} \\
        % \midrule
        $\log{S_0\omega_0^4}$ & GP amplitude & $-13.37^{+0.63}_{-0.61}$ & $\mathcal{N}(\log{\mathrm{var(flux)}}, 5^2)$ \\
        $\log{\omega_0}$ (\si{\per\second}) & GP frequency & $0.44^{+0.24}_{-0.28}$ & $\mathcal{N}(\log{2\pi / 5}, 5^2)$ \\
        $\log{\sigma^2}$ & White noise term & $-24.67^{+3.33}_{-5.35}$ & $\mathcal{N}(\log{\mathrm{var(flux)}}, 10^2)$ \\
        $\mathcal{F}$ & Flux scaling factor & $1.00005^{+0.00015}_{-0.00015}$ & $\mathcal{N}(\mathrm{median(flux)}, 5^2)$ \\[2pt]

        \bottomrule
    % \end{tabular*}
    \end{tabular}
    \label{table:results_gp}
\end{table*}

\begin{table*}%[htbp]
\renewcommand{\arraystretch}{1.2}
%    \sisetup{round-mode=places}
    \centering
    \contcaption{}
    \begin{tabular}{@{\extracolsep{\fill}}
    % \begin{tabular*}{\linewidth}{@{\extracolsep{\fill}}
            llcc}
        \toprule
        \toprule
        {\textbf{Parameter}} & {\textbf{Description}} & {\textbf{Value}} &  {\textbf{Prior}}\\% & {\textbf{Reference/data}}  \\
        \midrule
        % \midrule
        \multicolumn{4}{@{}l@{}}{$R$ \textit{band}} \\
        % \midrule
        $\log{S_0\omega_0^4}$ & GP amplitude & $2.87^{+1.82}_{-2.48}$ & $\mathcal{N}(-1, 10^2)$ \\
        $\log{\omega_0}$ (\si{\per\second}) & GP frequency & $5.10^{+2.29}_{-0.57}$ & $\mathcal{N}(\log{2\pi / 0.04}, 10^2)$ \\
        $\log{\sigma^2}$ & White noise term & $-10.98^{+0.25}_{-0.29}$ & $\mathcal{N}(\log{\mathrm{var(flux)}}, 10^2)$ \\
        $\mathcal{F}$ & Flux scaling factor & $0.89563^{+0.00080}_{-0.00069}$ & $\mathcal{N}(\mathrm{median(flux)}, 5^2)$ \\[2pt]
        \multicolumn{4}{@{}l@{}}{$z'$ \emph{band}} \\
        % \midrule
        $\log{S_0\omega_0^4}$ & GP amplitude & $-8.89^{+8.50}_{-3.46}$ & $\mathcal{N}(-1, 10^2)$ \\
        $\log{\omega_0}$ (\si{\per\second}) & GP frequency & $2.34^{+10.24}_{-2.53}$ & $\mathcal{N}(\log{2\pi / 0.04}, 10^2)$ \\
        $\log{\sigma^2}$ & White noise term & $-18.14^{+4.86}_{-7.64}$ & $\mathcal{N}(\log{\mathrm{var(flux)}}, 10^2)$ \\
        $\mathcal{F}$ & Flux scaling factor & $0.14724^{+0.00132}_{-0.00026}$ & $\mathcal{N}(\mathrm{median(flux)}, 5^2)$ \\[2pt]
        \multicolumn{4}{@{}l@{}}{$g'$ \textit{band}} \\
        % \midrule
        $\log{S_0\omega_0^4}$ & GP amplitude & $-3.14^{+1.27}_{-1.23}$ & $\mathcal{N}(-1, 10^2)$ \\
        $\log{\omega_0}$ (\si{\per\second}) & GP frequency & $2.84^{+0.93}_{-1.20}$ & $\mathcal{N}(\log{2\pi / 0.04}, 10^2)$ \\
        $\log{\sigma^2}$ & White noise term & $-19.04^{+4.66}_{-7.50}$ & $\mathcal{N}(\log{\mathrm{var(flux)}}, 10^2)$ \\
        $\mathcal{F}$ & Flux scaling factor & $0.16291^{+0.00197}_{-0.00266}$ & $\mathcal{N}(\mathrm{median(flux)}, 5^2)$ \\[2pt]
        \multicolumn{4}{@{}l@{}}{\emph{White light}} \\
        % \midrule
        $\log{S_0\omega_0^4}$ & GP amplitude & $3.10^{+1.11}_{-1.03}$ & $\mathcal{N}(-1, 10^2)$ \\
        $\log{\omega_0}$ (\si{\per\second}) & GP frequency & $3.97^{+0.42}_{-0.48}$ & $\mathcal{N}(\log{2\pi / 0.04}, 10^2)$ \\
        $\log{\sigma^2}$ & White noise term & $-10.90^{+0.21}_{-0.21}$ & $\mathcal{N}(\log{\mathrm{var(flux)}}, 10^2)$ \\
        $\mathcal{F}$ & Flux scaling factor & $1.36259^{+0.00500}_{-0.00481}$ & $\mathcal{N}(\mathrm{median(flux)}, 5^2)$ \\
        \multicolumn{4}{@{}l@{}}{\emph{SOPHIE}} \\
        % \midrule
        $\log{\sigma}$ (\si{\kilo\metre\per\second}) & White noise term &  $-5.44^{+1.34}_{-4.40}$ & $\mathcal{N}(\log{\mathrm{median(RV)}}, 5^2)$ \\
        $\gamma$ (\si{\kilo\metre\per\second}) & Systemic velocity & $-40.8197^{+0.0046}_{-0.0047}$ & $\mathcal{N}(\mathrm{median(RV)}, 0.5^2)$ \\
        \bottomrule
    % \end{tabular*}
    \end{tabular}
    % \label{table:results_gp2}
\end{table*}

\bsp	% typesetting comment
\label{lastpage}
\end{document}